\documentclass[10pt,letterpaper]{article}
\usepackage[top=0.85in,left=2.75in,footskip=0.75in]{geometry}

\usepackage{amsmath,amssymb}

\usepackage{changepage}

\usepackage{textcomp,marvosym}

\usepackage{cite}

\usepackage{nameref,hyperref}

\usepackage[right]{lineno}

\usepackage[nopatch=eqnum]{microtype}
\DisableLigatures[f]{encoding = *, family = * }

\usepackage[table]{xcolor}

\usepackage{array}

\newcolumntype{+}{!{\vrule width 2pt}}

\newlength\savedwidth

\raggedright
\usepackage[aboveskip=1pt,labelfont=bf,labelsep=period,justification=raggedright,singlelinecheck=off]{caption}

\makeatletter
\renewcommand{\@biblabel}[1]{\quad#1.}
\makeatother

\usepackage{lastpage,fancyhdr,graphicx}
\usepackage{epstopdf}
\fancyheadoffset[L]{2.25in}
\fancyfootoffset[L]{2.25in}
\usepackage{subfiles}
\usepackage{framed}
\usepackage{siunitx}
\usepackage{amsmath}
\usepackage{pgfplots}
\pgfplotsset{compat=1.18} 
\usetikzlibrary{arrows, arrows.meta}
\usepackage{listings}
\usepackage{subcaption}
\usepackage{graphicx}
\usepackage{tikz}
\usetikzlibrary{positioning}
\usepackage{calc}
\usetikzlibrary{calc}
\usepackage{cleveref}
\usepackage{float}
\usepackage{array,tabularx,booktabs}
\usepackage[table]{xcolor}
\usepackage{arydshln}
\usepackage{ragged2e}
\usepackage{multirow}
\usepackage{multicol}
\usepackage{makecell} 
\usepackage[table]{xcolor}
\usepackage{eurosym}
\usepackage{enumitem}
\usepackage{derivative}

\newcolumntype{P}[1]{>{\centering\arraybackslash}m{#1}}

\newcolumntype{Y}{>{\Centering\arraybackslash}X}
\newcolumntype{C}[1]{>{\centering\arraybackslash}p{#1}}

\definecolor{heat0}{HTML}{FFFFFE}
\definecolor{heat1}{HTML}{d6eacf}
\definecolor{heat2}{HTML}{add4a0}
\definecolor{heat3}{HTML}{83bf72}
\definecolor{heat4}{HTML}{56a943}
\definecolor{heat5}{HTML}{119300}

\newcommand{\heatcell}[1]{%
  \ifcase#1
    \cellcolor{heat0}#1
  \or
    \cellcolor{heat1}#1
  \or
    \cellcolor{heat2}#1
  \or
    \cellcolor{heat3}#1
  \or
    \cellcolor{heat4}#1
  \or
    \cellcolor{heat5}#1
  \else
    #1%
  \fi
}

\DeclareSIUnit\angstrom{\text {Å}}

\begin{document}
\vspace*{0.2in}

\begin{flushleft}
{\Large
\textbf{Sensing the properties of virtual objects without physical feedback} % Please use "sentence case" for title and headings (capitalize only the first word in a title (or heading), the first word in a subtitle (or subheading), and any proper nouns).
}
\newline
\\
Rhoslyn Roebuck Williams\textsuperscript{1*},
Harry J. Stroud\textsuperscript{1},
Luis E. Toledo\textsuperscript{1\Yinyang},
Mark D. Wonnacott\textsuperscript{1\Yinyang},
David R. Glowacki\textsuperscript{1*}
\\
\bigskip
\textbf{1} Intangible Realities Laboratory (IRL), Centro Singular de Investigación en Tecnoloxías Intelixentes (CiTIUS),
University of Santiago de Compostela, Santiago de Compostela, Spain
\\
% \textbf{2} Affiliation Dept/Program/Center, Institution Name, City, State, Country
% \\
% \textbf{3} Affiliation Dept/Program/Center, Institution Name, City, State, Country
% \\
\bigskip

% Insert additional author notes using the symbols described below. Insert symbol callouts after author names as necessary.
% 
% Remove or comment out the author notes below if they aren't used.
%
% Primary Equal Contribution Note
\Yinyang These authors contributed equally to this work.

% Additional Equal Contribution Note
% Also use this double-dagger symbol for special authorship notes, such as senior authorship.
% \ddag These authors also contributed equally to this work.

% \textcurrency Current Address: Centro Singular de Investigación en Tecnoloxías Intelixentes (CiTIUS),
% University of Santiago de Compostela, 15782, Santiago de Compostela, Spain

% \textcurrency Current Address: Dept/Program/Center, Institution Name, City, State, Country % change symbol to "\textcurrency a" if more than one current address note
% \textcurrency b Insert second current address 
% \textcurrency c Insert third current address

% \dag Deceased

% Group/Consortium Author Note
% \textpilcrow Membership list can be found in the Acknowledgments section.

% Use the asterisk to denote corresponding authorship and provide email address in note below.
* Corresponding authors: rhoslyn.roebuckw@usc.es (RRW); drglowacki@gmail.com (DRG)

\end{flushleft}

% For PLOS Medicine research article authors, please structure your abstract
% with "Background", "Method and Findings" and "Conclusion" sections per
% journal requirements.

% For PLOS Neglected Tropical Diseases research article authors, please
% structure your abstract with "Background", "Methodology", "Findings", and
% "Conclusion" sections per journal requirements.
%

% --------------------------------------------------- %

\section*{Abstract}

People who have interacted with simulated worlds and simulated objects in extended reality (XR) often have a sense that they can `feel' the objects being simulated despite them not being physical.
Our sense of touch is essential for how we `feel' the physical world, however, there is an open question as to what it means to `feel' virtual objects when interacting with them in immersive digital environments.
In prior research, we have reported that participants often describe a subjective experience of `feeling' the properties of simulated molecular objects while using interactive molecular dynamics in extended reality (iMD-XR), a field-based interaction paradigm for manipulating real-time simulations of molecular objects without haptic feedback.
To better understand these subjective reports of `feeling', we used a psychophysics approach to quantify the threshold at which participants perceive differences in the rigidity of simulated molecular objects (C$_{60}$ molecules) in iMD-XR.
To evaluate this, we carried out experiments to compare the just-noticeable differences (JNDs) in two conditions: (1) via direct interaction with a real-time C$_{60}$ simulation, and (2) via observation-only---i.e. watching another person interacting with the simulations.
Our findings show that direct interaction enabled participants to perceive more subtle rigidity differences of \SI{11.5}{\percent}, compared to \SI{18.5}{\percent} for observation-only.
Furthermore, participants who undertook interaction first were better able to distinguish rigidity differences in the subsequent observation-only condition, suggesting that interaction trained participants to better perceive differences in molecular properties.
These findings demonstrate a novel and flexible approach for sensing the properties of virtual objects in XR, and offer new insights into iMD-XR's potential in molecular research and education.

% --------------------------------------------------- %
% Please keep the Author Summary between 150 and 200 words. Use first person.
% PLOS ONE, PLOS Biology, PLOS Global Public Health, PLOS Mental Health, and PLOS Water authors please skip this step. Author Summary is not valid for submissions to these journals.

% For PLOS Medicine authors, please structure your author summary with answers to the following questions:
% Why was this study done?
% What did the researchers do and find?
% What do these findings mean?
%
% \section*{Author summary}
% Lorem ipsum dolor sit amet, consectetur adipiscing elit. Curabitur eget porta erat. Morbi consectetur est vel gravida pretium. Suspendisse ut dui eu ante cursus gravida non sed sem. Nullam sapien tellus, commodo id velit id, eleifend volutpat quam. Phasellus mauris velit, dapibus finibus elementum vel, pulvinar non tellus. Nunc pellentesque pretium diam, quis maximus dolor faucibus id. Nunc convallis sodales ante, ut ullamcorper est egestas vitae. Nam sit amet enim ultrices, ultrices elit pulvinar, volutpat risus.

% \linenumbers

% Use "Eq" instead of "Equation" for equation citations.

% --------------------------------------------------- %

\section*{Introduction}
\label{sec:introduction}

% \begin{quote}
%     ``\textit{VR devices aren't illusion machines; they're reality machines.}''
% \end{quote}
% % David Chalmers quote (Page 206)

% One of the most remarkable things about XR is the ability to create experiences that are otherwise impossible in the physical world. 
% 

We use our sense of touch every day to interact with and explore physical objects, yet generating this experience when interacting with virtual objects is challenging.
In some cases, haptic feedback---where touch sensations are generated via physical stimulation---can be used to provide physical feedback when interacting with virtual objects.
However, the implementation of such feedback is difficult due to the cost and availability of hardware, and the need for bespoke integration.
Moreover, haptic technology faces a more fundamental limitation: namely that current solutions (including, for example, vibrotactile feedback, surface haptics, force feedback, ultrasonics and thermal feedback) are not generalisable and cannot reproduce the full range of tactile sensations that we experience in the physical world \cite{slater_enhancing_2016}.

% enhance the sense of realism in virtual environments, however, their implementation is limited by the cost and availability of hardware, and the need for bespoke integration. 
% We use our sense of touch every day to interact with and explore physical objects.
% However, creating this experience in virtual environments is challenging.
% In some cases, haptic feedback---where touch sensations are generated via physical stimulation---can be used to enhance the sense of realism in virtual environments, however, implementing haptics remains challenging due to the cost and availability of hardware, and the need for bespoke integration. 
% As such, haptic feedback---where touch sensations are generated via physical stimulation---in some cases offers a solution to enhance the sense of realism in virtual environments.
% However, implementing haptic feedback remains challenging due to the cost and availability of hardware, and the need for bespoke integration.

Given these limitations with haptic technology, there is an open research question: to what extent are participants in XR environments able to detect the properties of interactive virtual objects \textit{without} physical haptic feedback?
One approach is to introduce controlled conflicts between the visual and motor cues---termed visuomotor incongruence---to induce visuo-haptic illusions, i.e. `haptic-like' sensations.
For example, Samad et al. demonstrated that offsetting a participant's virtual hand position from their physical hand when lifting a virtual object in VR could induce a sensation of weight~\cite{samad_pseudo-haptic_2019}.
Owing to the fact that no physical stimulation is involved, these are termed \textit{pseudo-haptic} approaches (see~\cite{kurzweg_survey_2024, ujitoko_survey_2021, collins_pseudo-haptics_2019, lecuyer_simulating_2009} for reviews on pseudo-haptics).

A range of pseudo-haptic sensations have been reported in VR, including sensations of weight~\cite{samad_pseudo-haptic_2019, stellmacher_continuous_2023}, stiffness \cite{ariza_nunez_holitouch_2022, weiss_using_2023} and resistance \cite{pusch_hemp-hand-displacement-based_2008,stepanova_ethereal_2026}.
% These studies employ psychophysical techniques to quantify detection thresholds, whilst self-report measures provide complementary insights into participants’ subjective experiences, together informing the design of interactive virtual environments.
Though effective, deliberate introduction of visuomotor incongruence risks disrupting one's sense of presence within VR and has been shown to reduce participants' proprioceptive accuracy with prolonged exposure \cite{feick_delusionized_2025}.
Furthermore, this approach requires finding a sensory discrepancy that can reliably induce a given haptic-like sensation through cross-modal interaction, thereby limiting its flexibility and broader applicability.

\begin{figure}
    \centering
    \includegraphics[width=\linewidth]{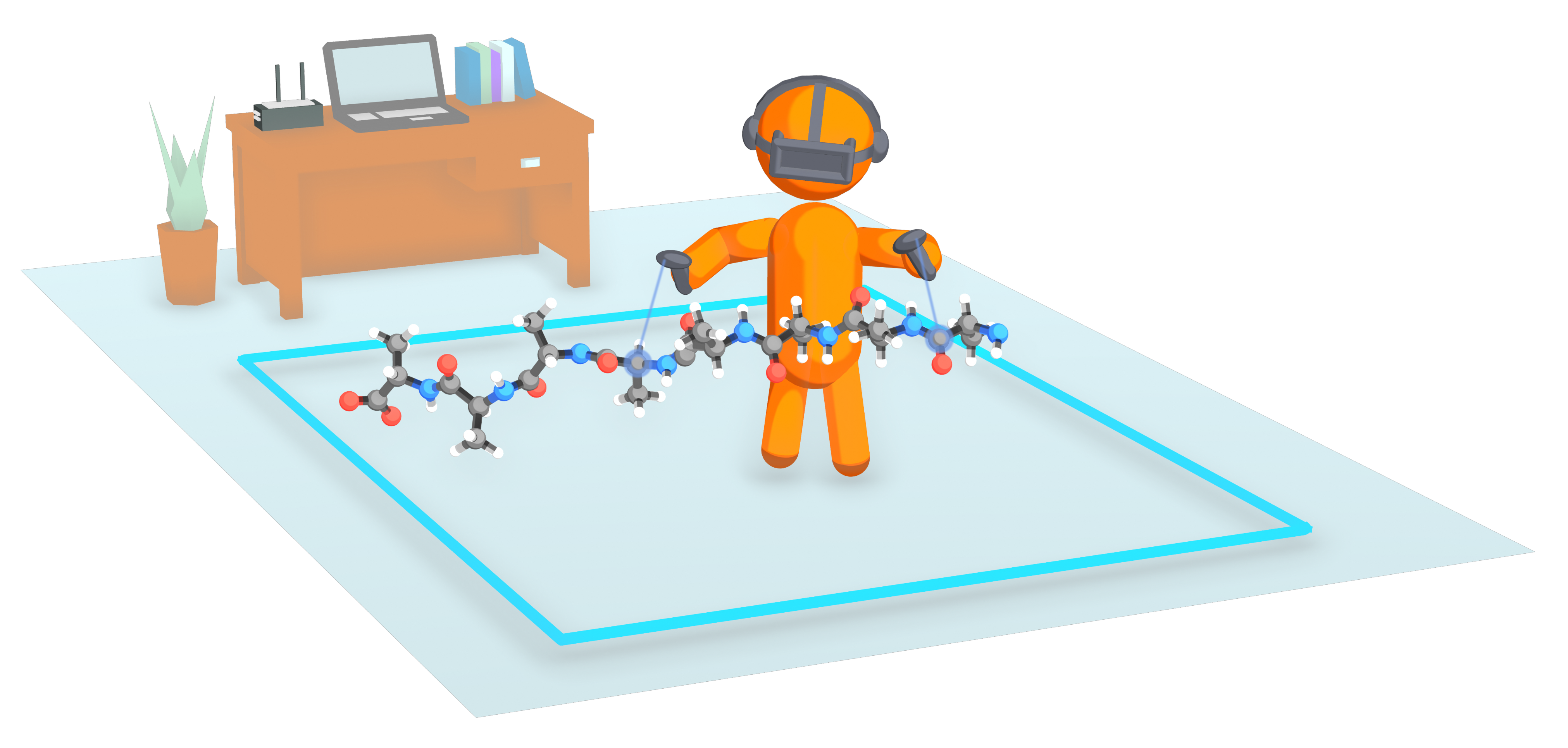}
    \caption{
    \textbf{A participant interacts with a virtual polypeptide in iMD-XR.}
    }
    \label{fig:iMD-XR_illustration}
\end{figure}

Over the last several years, we have taken a different approach to studying the perception of virtual objects in the absence of haptics, which we have called ‘Subtle Sensing’ \cite{roebuck_williams_subtle_2020}.
% Subtle Sensing is an alternative interaction paradigm based on the observation that XR participants often indicate they can `sense' the dynamic properties of flexible virtual objects, despite the absence of both physical feedback and sensory incongruence.
The Subtle Sensing paradigm has its origins in our work developing a software framework for carrying out interactive molecular dynamics in extended reality (iMD-XR).
With iMD-XR, participants enter immersive XR environments where they can interact with live molecular simulations, applying forces using natural hand motions and experiencing the dynamic response of the system in real time, as illustrated in \Cref{fig:iMD-XR_illustration}.
Prior studies have shown that iMD-XR participants can perceive differences in molecular properties despite the absence of physical feedback or sensory
incongruence. 
For example, O'Connor et al. \cite{oconnor_interactive_2019} reported their `Burke Perception Experiment', during which Prof. Kieron Burke described how a polypeptide ``\textit{feels so much different}'' than a carbon nanotube when interacting with the simulations in the iMD-XR environment.
% Similarly to pseudo-haptic approaches, iMD-XR participants have been shown to perceive the properties of interactive molecular objects without physical feedback. 
% However, unlike pseudo-haptic techniques, this perception arises without the introduction of visuomotor incongruence. 
Roebuck Williams et al. \cite{roebuck_williams_subtle_2020} followed this up with a pilot study showing that participants could perceive rigidity differences of C$_{60}$ molecules and string-like `polypeptide' molecules.
A further study extended this work as a proof-of-concept for quantifying perception thresholds of molecular properties using a two-alternative forced-choice (2AFC) psychophysics approach \cite{roebuck_williams_measuring_2024}.

% PARAGRAPH HERE -about why molecules are ideal for studying perception
`Subtle Sensing' is focused on the study of virtual objects whose physics are flexible and dynamic, which requires more complex physics modelling and alternative rendering approaches compared to the rigid-body dynamics typically used in XR environments.
In this regard, molecules offer good candidates for studying the dynamics and perception of flexible objects, given that there are well-established methods for simulating and visualising molecular systems, along with rendering techniques that are specifically designed to support perception of molecular structure and dynamics. 
A range of software packages are available for preparing and simulating molecular objects, rigorously tuning their properties, and performing detailed analyses of their movements.
Open-source platforms such as NanoVer \cite{stroud_nanover_2025,wonnacott_nanover_2026} and UnityMol \cite{doutreligne_unitymol_2014} enable direct interaction with these virtual objects in XR.

Molecules provide a particularly interesting application for the Subtle Sensing paradigm, and more generally for investigating alternatives to physical haptic feedback in XR.
First and foremost is the fact that nobody knows what a molecule should ‘feel’ like, i.e. there is no ground truth given that we cannot experience \textit{touching} physical molecules. 
Furthermore, the nanoscale world follows complex and unfamiliar physical laws that are unlike everyday physical objects \cite{sutherland_ultimate_1965}.
The fact that most people have no prior experience of interacting with a molecular object makes molecular systems an intriguing test case for interaction in XR. 
When we interact with virtual objects that replicate familiar physical objects, we expect a given behaviour based on our prior experience. 
In contrast, we do not have the corresponding familiarity with molecular objects, since they have no physical analogue.
Using the Bayesian language of predictive processing models of cognition, we have relatively ill-defined `priors' for interacting with molecules in XR \cite{laukkonen_beautiful_2025}, an observation which similarly extends to other classes of virtual experience developed in our lab over the years \cite{glowacki_vr_2024}.
When encountering objects for which we have ill-defined priors, we engage in a process of `active inference', constructing predictive models of behaviour that minimise prediction error (or surprise) in future encounters.
This suggests that there may be a window during which participants are more receptive to `bottom-up' sensory data and fully engage in the practice of Subtle Sensing to build a novel world model through a process of active inference \cite{laukkonen_beautiful_2025}.

\begin{figure}
    \centering
    \includegraphics[width=\linewidth]{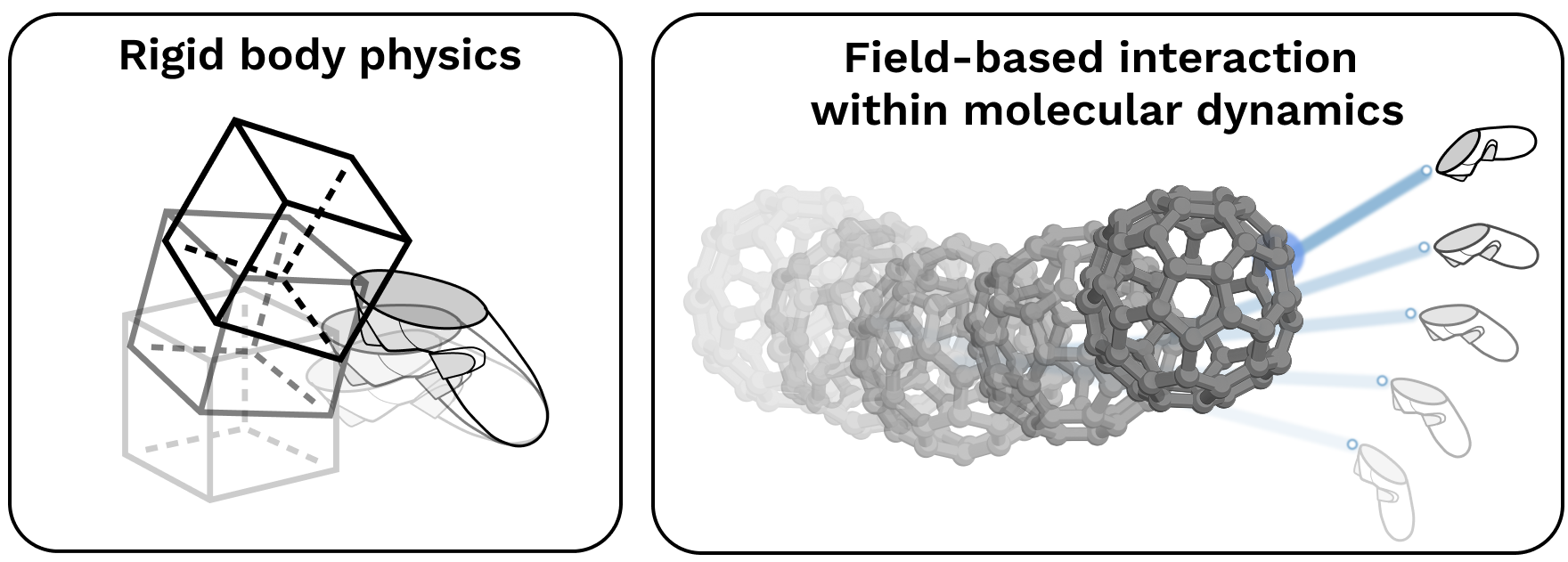}
    \caption{
    \textbf{Illustration of (a) interaction with a virtual cube simulated with rigid body physics and (b) field-based interaction with a virtual molecule simulated using molecular dynamics.}
    When interacting with the virtual molecule, the player uses the controller to apply a force to a single atom.
    Videos of field-based interaction within iMD-XR are provided in the Supporting Information. 
    }
    \label{fig:interaction_paradigms}
\end{figure}

Our previous Subtle Sensing work has suggested that participants can indeed sense molecular rigidities in iMD-XR. 
This raises an important question: how can we perceive these properties in the absence of physical haptic feedback? 
As illustrated in \Cref{fig:interaction_paradigms}, virtual molecules do not obey the same rigid body dynamics typical of virtual objects in XR. 
Instead, iMD-XR participants interact with virtual molecules by perturbing the force field of the molecular simulation, a paradigm we refer to as \textit{field-based interaction}.
Participants can perturb this force field from any position in the virtual space.
% , while colocation between their physical and virtual bodies is maintained. 
During interaction, the interaction origin (the participant's hand) and the object of interaction (the molecule) are coupled, and the force applied to the molecule depends on their relative positions. 
However, the molecule's response is also governed by other forces within the system, such that the resulting dynamics reflect both the applied force and the intrinsic forces acting within the molecular system. 
By paying attention to how the molecule responds to their perturbations, participants can infer information about these intrinsic forces. 
In this way, they are able to \textit{sense the properties} of the molecular simulation \textit{without physical feedback}.

Many participants describe this experience of sensing molecular properties as `feeling' them (e.g. the aforementioned `Burke Perception Experiment').
However, there remains an open question about what it means to `feel' simulated molecules, or to `feel' virtual objects more generally.
A recent study involving EEG measurements showed that virtual touch in XR---where participants were touched by a virtual robot arm without physical feedback---was processed similarly to physical touch, activating sensorimotor regions in the brain characteristic of physical touch despite the absence of direct physical stimulation \cite{savalle_does_2026}.
This finding offers evidence that we can `feel' virtual objects to some extent, even without physical feedback.

\subsection*{Research goals}

In this article, we report two user studies that quantitatively measured the threshold at which participants perceive rigidity differences of interactive molecular simulations in iMD-XR. 
Our goal was to determine the extent to which perception of these properties is influenced by direct interaction versus observation of another participant's interactions.

This work builds on earlier studies exploring the role of interaction in perceiving molecular properties \cite{roebuck_williams_subtle_2020, roebuck_williams_measuring_2024}.
The current study overcomes several deficiencies in the prior studies, including the lack of automation and statistical analysis, reliance on participant memory, and the absence of a protocol for quantitative comparison between experimental conditions. 
% [something about statistics (power)? Improved sampling?]
The studies outlined herein incorporate: (a) a custom iMD-XR game to control experimental conditions across participants, (b) on-the-fly data collection within the virtual environment, and (c) a 2AFC approach to enable quantitative estimation of perceptual thresholds with improved sampling of the psychometric function.

Our results show that participants could distinguish differences in the bond-angle rigidity of C$_{60}$ molecules both when interacting directly and when observing an expert’s interactions in XR.
We found that direct interaction led to better performance than observing-only, and that prior interaction improved subsequent performance during the observing-only condition. 
% By comparing our estimated rigidity perception thresholds with bond-angle rigidity of another molecular system, we demonstrate that such perceptual sensitivity may be useful in distinguishing rigidities of other chemical systems.
Our findings illustrate the potential of iMD-XR in molecular research and education, and demonstrate Subtle Sensing as a novel approach for conveying the dynamic properties of interactive virtual objects without physical feedback.

% --------------------------------------------------- %
\section*{Methods}
\label{sec:methods}

\subsection*{Overview}

Two user studies were conducted using two-alternative forced-choice (2AFC) procedures to investigate perception of bond-angle rigidity in interactive C$_{60}$ simulations in iMD-XR. 
Ethical approval for the studies was granted by the Bioethics Committee at the University of Santiago de Compostela, Spain (USC 51/2022).

\paragraph*{Study One.}
Study One ($N=28$) examined two conditions within the 2AFC task: (a) the \textit{Interacting} condition, in which participants directly interacted with the molecular simulations in XR, and (b) the \textit{Observing} condition, in which participants watched recordings of an experimenter interacting with the simulations from within XR. 
This study sampled increasing rigidities of the C$_{60}$ molecules (buckyballs).
A within-subjects design was used, and the order of conditions was counterbalanced across participants using random assignment.
Data were collected in three phases and participants in the first ($n=7$) and second ($n=7$) phases were entered into prize draws (see \nameref{S1_Table} for details).
Participants in the third phase did not receive any compensation.

\paragraph*{Study Two.}
Study Two ($N=14$) was a smaller-scale, complementary study to Study One with the primary aim of improving the sampling in the 2AFC task.
% In particular, we sampled a different range of rigidity values to Study One and increased the number of trials per participant in order to obtain a more precise estimate of the threshold at which participants could reliably discriminate between molecules.
Where Study One sampled increasing rigidities, this study sampled decreasing rigidities to examine whether perceptual sensitivity was comparable across both regimes.
Furthermore, the number of trials per participant was increased to improve the accuracy of the estimated perceptual threshold.
This was achieved by removing the Observing condition and doubling the number of Interacting trials.
As the aim was to calculate accurate threshold values, we included expert iMD-XR users (including three authors of this work), to reduce the impact of learning effects on threshold estimation.
The lead author did not participate in the study, and was responsible for conducting all experiments across both studies.
Only one experimental condition was implemented, therefore no counterbalancing was required. 
Non-expert participants were entered into a prize draw for participation, and expert participants received no compensation (see \nameref{S1_Table}).

\subsection*{System design and implementation}

\subsubsection*{SubtleGame}

\begin{figure}
    \centering
    \includegraphics[width=0.95\linewidth]{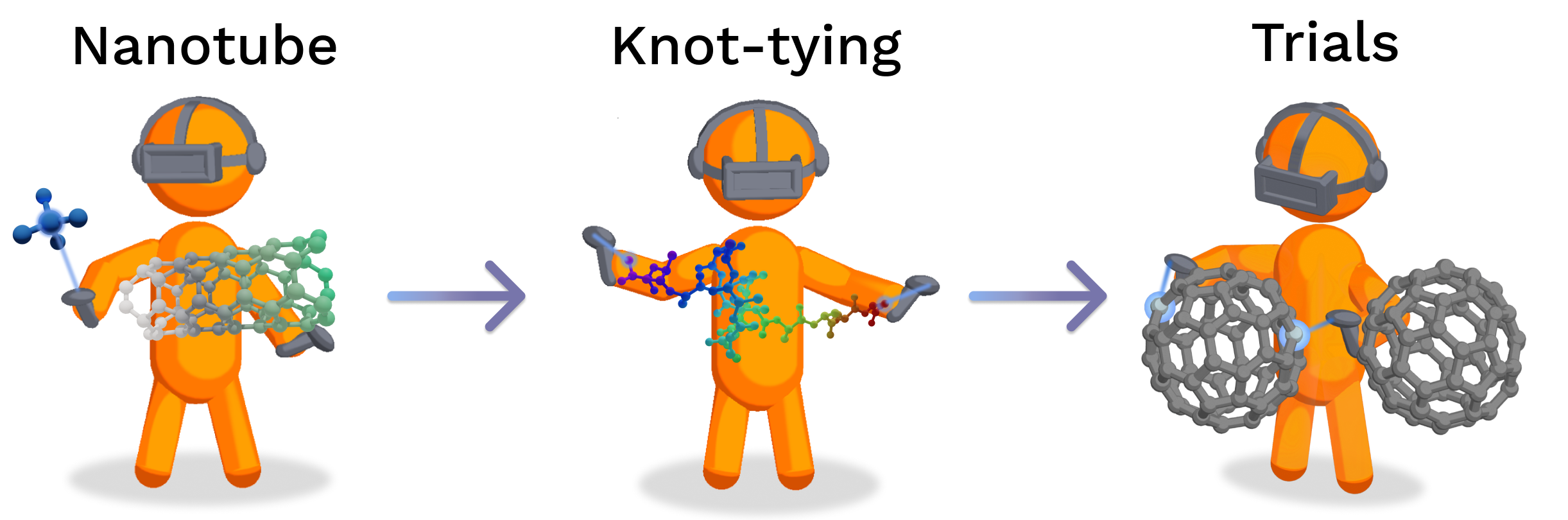}
    \vspace{0.5em}
    \caption{
    \textbf{Illustrations of the Nanotube, Knot-tying and Trials tasks within the SubtleGame XR application.}
    }
    \label{fig:illustrations_games_in_subtlegame}
\end{figure}

The studies were conducted using \textit{SubtleGame}, a single-player XR application developed in our lab that utilises NanoVer, a software framework for iMD-XR \cite{stroud_nanover_2025,wonnacott_nanover_2026}.
Within SubtleGame, players engage in three tasks: the Nanotube, Knot-tying, and Trials tasks (\Cref{fig:illustrations_games_in_subtlegame}).
In the Nanotube task, players guide methane through the centre of a carbon nanotube.
In the Knot-tying task, players tie a simple knot in the 17-alanine polypeptide.
The Nanotube and Knot-tying tasks were included in this study to (a) provide training in interacting with molecular simulations, (b) break up the monotony of the 2AFC tasks, and (c) contribute data to other ongoing research projects. 
Data from these tasks are not included in the present work.

A spherical Gaussian interaction potential was employed to model user-applied forces (see \Cref{eq:gaussian_potential} and \Cref{fig:spherical_gaussian_potential} in the Technical details section).
In this paradigm, the magnitude of the applied force increases with controller-atom separation up to a predefined interaction radius of \SI{1}{\nano\metre}, beyond which it decays smoothly towards zero.
For reference, the diameter of a buckyball is approximately \SI{0.7}{\nano\metre}.
Consequently, participants cannot exert forces on atoms from large distances.
This potential was selected to limit the amount of energy that could be introduced into the system, thereby maintaining simulation stability and encouraging participants to maintain close proximity to the molecules.

\subsubsection*{The Trials task}

\begin{figure}
    \centering
    \includegraphics[width=\linewidth]{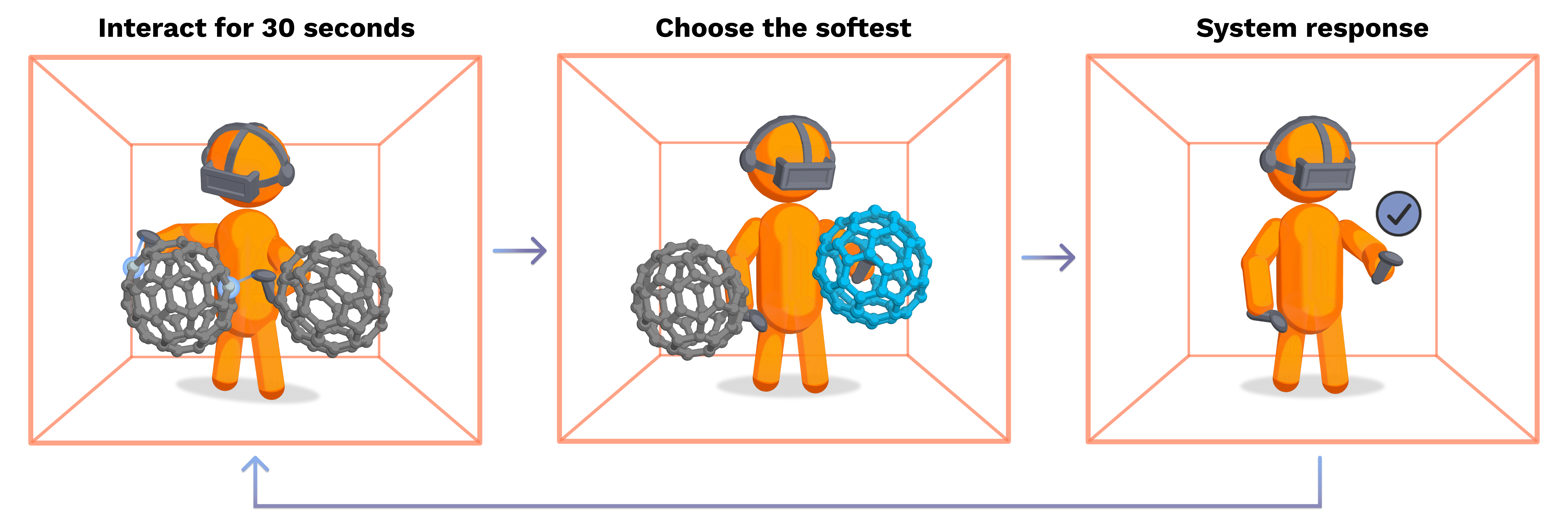}
    \caption{
    \textbf{The 2AFC procedure within SubtleGame.}
    }
    \label{fig:2afc_trial_procedure}
\end{figure}

The Trials task constituted a gamified 2AFC psychophysics experiment using the method of constant stimuli, which was adapted from that used by Roebuck Williams et al. \cite{roebuck_williams_measuring_2024}.
This procedure is illustrated in \Cref{fig:2afc_trial_procedure}, and a video is provided in the Supporting Information.
During each trial, participants were presented with two buckyballs for a maximum duration of 30 seconds. 
One buckyball served as the reference and was identical in every trial, while the rigidity of the other was modified to varying degrees (more details below). 
The two buckyballs appeared in front of the participant, positioned side-by-side in a randomised order.
After 30 seconds had elapsed---or once the participant had pressed the ‘Answer Now’ button---the simulation paused, and the participant indicated the softest buckyball by holding their controller inside the chosen buckyball for two seconds. 
A tick or cross symbol appeared above the participant’s controller to indicate the correctness of the response, and a summary of their responses was displayed on an information panel on the side of the simulation box. 
Upon completion of the task, the menu displayed the participant’s total score as a percentage.

At the start of the Trials task for each experimental condition, participants completed two training trials (each lasting up to one minute) featuring buckyballs with large differences in rigidity.
This allowed participants to familiarise themselves with the game mechanics and understand how to compare the rigidities of the buckyballs.

\subsubsection*{C$_{60}$ simulations}

The molecular dynamics (MD) simulations were propagated in real time from a set of specified initial conditions using Newton's equations of motion. % $\{ \mathbf{q} (t_0), \mathbf{v} (t_0) \}$
These equations were integrated iteratively to obtain the atomic positions $\mathbf{q}$ and velocities $\mathbf{v}$ as functions of time $t$ as follows:
\begin{align}
    \mathbf{v}_i  \Big( t + \frac{\Delta t}{2} \Big) 
    &= 
    \mathbf{v}_i  \Big( t - \frac{\Delta t}{2} \Big) + \frac{\mathbf{f} _i (t) }{m_i} \Delta t
    \label{eq:eom_velocities}\\[0.5em]
    \mathbf{q}_i  \Big( t + \frac{\Delta t}{2} \Big) 
    &= 
    \mathbf{q}_i  \big( t \big) + \mathbf{v}_i  \Big( t + \frac{\Delta t}{2} \Big) \frac{\Delta t}{2}
    \label{eq:eom_positions}
\end{align}
where $i$ is the atom index, 
$\mathbf{q}_i$ is the position of atom $i$,
$\mathbf{v}_i$ is its velocity,
$\mathbf{f}_i$ is the force acting on it,
$m_i$ is its mass,
and $\Delta t$ is the integration timestep.
To maintain simulation stability during user interaction, damped velocities $\mathbf{v}'_i$ were calculated using a so-called \textit{Langevin thermostat} (see the SI for further information), which combines a friction force with random noise to ensure that the time-dependent atomic dynamics fluctuate around an average that is characteristic of the specified temperature (\SI{300}{\kelvin}).
The on-step positions were then calculated from the damped velocities $\mathbf{v}'_i(t + \frac{\Delta t}{2})$ as follows:
\begin{align}
    \mathbf{q}_i (t + \Delta t) 
    &= 
    \mathbf{q}_i \Big( t + \frac{\Delta t}{2} \Big) + \mathbf{v}'_i  \Big( t + \frac{\Delta t}{2} \Big) \frac{\Delta t}{2}
\end{align}
These time series of atomic positions and velocities define a sequence of `snapshots' of molecular configurations known as a \textit{trajectory}.

% These time series of atomic positions and velocities form a series of ‘snapshots’ defining a \textit{trajectory}.
% These time series of $\mathbf{q}(t)$ and $\mathbf{v}(t)$ form a series of ‘snapshots’ called a trajectory.

The forces $\mathbf{f}_i$ in \Cref{eq:eom_velocities} comprise the components of the force vector $\mathbf{F} (\mathbf{q})$, which are calculated as the negative gradient of the potential energy function $V(\mathbf{q})$ with respect to the vector of atomic coordinates $\mathbf{q}$:
\begin{align}
    \mathbf{F} (\mathbf{q})
     = - \odv{}{\mathbf{q}} \Big( V(\mathbf{q}) \Big)
\end{align}

In iMD-XR, the potential energy function $V(\mathbf{q})$ is defined as the sum of the internal potential energy of the C$_{60}$ molecules $V_\text{int}(\mathbf{q})$ and the external interaction potential $V_\text{user}(\mathbf{q})$. 
$V_\text{int}(\mathbf{q})$ is defined by three energy terms: 
an anharmonic angle-bending potential $V_\text{angle}(\mathbf{q})$, 
an anharmonic bond-stretching potential $V_\text{bond}(\mathbf{q})$, 
and a Lennard–Jones potential $V_\text{nonbonded}(\mathbf{q})$ that models nonbonded interactions.
$V_\text{user}(\mathbf{q})$ depends on the distance between the participant’s XR controllers and the atoms that are the target of the user's interactions. 
The rigidity of the buckyballs was modified by scaling the anharmonic angle-bending term by a scaling factor $x$. 
Thus, the total force vector $\mathbf{F}$ is calculated in terms of the total potential energy $V(\mathbf{q})$ as follows:
\begin{align}
\mathbf{F} (\mathbf{q})
     &= - \odv{}{\mathbf{q}} \Big( V_\text{int}(\mathbf{q}) + V_\text{user}(\mathbf{q}) \Big)
\\
    &= - \odv{}{\mathbf{q}}
     \Big(
        x \, V_\text{angle} (\mathbf{q})
        + V_\text{bond} (\mathbf{q})
        + V_\text{nonbonded} (\mathbf{q})
        + V_\text{user} (\mathbf{q})
     \Big)
 \label{eq:buckyball_total_force_a}
\\[0.5em]
    &= x \mathbf{F}_\text{angle} (\mathbf{q})
         + \mathbf{F}_\text{bond} (\mathbf{q})
         + \mathbf{F}_\text{nonbonded} (\mathbf{q})
         + \mathbf{F}_\text{user} (\mathbf{q})
\label{eq:buckyball_total_force_b}
\end{align}

% The potential energy function describing each contribution to the angle-bending term has a standard anharmonic form defined in terms of a $\mathbf{q}$-dependent angle $\theta (\mathbf{q})$ as:

The angle-bending potential in \Cref{eq:buckyball_total_force_a} is the sum of terms across the set of bonded C--C--C angles: 
\begin{align}
    V_\text{angle}(\mathbf{q}) 
    = \sum _\alpha V_{\text{angle,\,}\alpha} (\mathbf{q})
\end{align}
where $\alpha$ is the bond angle index. 
Each term takes a standard anharmonic form defined in terms of the angle $\theta$ formed by the C--C--C bonds (which is a function of $\mathbf{q}$):
\begin{align}
    \begin{split}
    % V_{\text{angle,}\alpha} (\mathbf{q}) 
    V_{\text{angle,\,}\alpha} (\mathbf{q}) = 
    \frac{1}{2}k ( \Delta \theta) ^2 \, 
    \Big(  &
    1 
    - 0.014 \Delta \theta 
    + \SI{5.6e-5}{} \, ( \Delta \theta ) ^2 \\ 
    &- \ \SI{1e-6}{} \, ( \Delta \theta ) ^3 
    + \ \SI{2.2e-8}{} \, ( \Delta \theta ) ^4 
    \Big)
    \end{split}
\label{eq:buckyball_angle_potential}
\end{align}
where $k$ is the angle force constant and $\Delta \theta  = \theta - \theta _{\text{eq}}$, where $\theta _{\text{eq}}$ is the equilibrium C--C--C bond angle. 

When a participant deforms a buckyball, the internal forces oppose the applied force and act to restore the equilibrium geometry.
According to \Cref{eq:buckyball_total_force_a,eq:buckyball_total_force_b}, the magnitude of these restoring forces depends on the rigidity scaling factor, $x$.
Participants can therefore probe the system by applying perturbing forces in order to perceive differences in rigidity.
Scaling factors $x>1$ increase resistance to deformation (analogous to an inflated ball), whereas scaling factors $x<1$ decrease resistance to deformation (analogous to a deflated ball).
Videos showing buckyballs with different rigidities are provided in the SI. 
% \nameref{S1_Video}--\nameref{S7_video}. \nameref{S2_Text}.

The range of scaling factors used in Studies One and Two was chosen based on the pilot study by Roebuck Williams et al. \cite{roebuck_williams_measuring_2024}. 
Scaling factors were spaced logarithmically according to a geometric series, with the same relative values used in Study One ($x_{\texttt{hard}} = 1.036$ to $1.375$) and Study Two ($x_{\texttt{soft}} = 0.727$ to $0.965$).
The training trials also used the same relative scaling factors for Study One ($x_{\texttt{hard}} = 1.7$) and Study Two ($x_{\texttt{soft}} = 0.588$). 
Participants completed a total of 50 study trials: 5 scaling factors $\times$ 5 trials $\times$ 2 conditions (Study One), and 5 scaling factors $\times$ 10 trials $\times$ 1 condition (Study Two).
Participants also performed four training trials, which were not included in the analysis.

The force field for a C$_{60}$ molecule is comprised of 90 individual angle force constants. 
'Rigidiy' on the other hand is a global property of the C$_{60}$ molecule.
To quantify how the scaling factors applied to individual angle force constants ($x$ in \Cref{eq:buckyball_total_force_a,eq:buckyball_total_force_b}) affected the collective rigidity of a C$_{60}$ molecule, we measured the maximum deformation of the molecule across the experimental scaling factors, as described in \nameref{S4_Text}.
The results indicate that the maximum deformation of a C$_{60}$ molecule scales approximately linearly with the logarithm of the scaling factor.

\subsubsection*{Interacting condition}

In the Interacting condition, participants interacted directly with the buckyballs using their XR controllers.
Two simulations were prepared for each scaling factor to counterbalance the left--right positions of the buckyballs relative to the participant.
This was done to reduce the impact of bias, e.g. participants systematically selecting the same buckyball when uncertain.

\subsubsection*{Observing condition}
% Within the Trials task, the Interacting and Observing conditions differed in how participants were able to interact with the molecular simulations.

% To minimise differences between the experimental conditions, during the observing-only condition participants watched recordings of an iMD-XR expert interacting with the molecular simulations from within the virtual environment.
% During these recordings, the iMD-XR expert applied interactions in a predefined and consistent manner (to within natural human variation), and was blinded to order of presentation of the simulations.

In the Observing condition, participants watched a recording of an experimenter interacting with the buckyballs in XR. 
They could move freely within the virtual environment to observe the buckyballs, as well as the experimenter’s avatar and interactions with the buckyballs, however, participants were unable to interact with the simulation themselves.

\begin{figure}
\centering
\includegraphics[width=\linewidth]{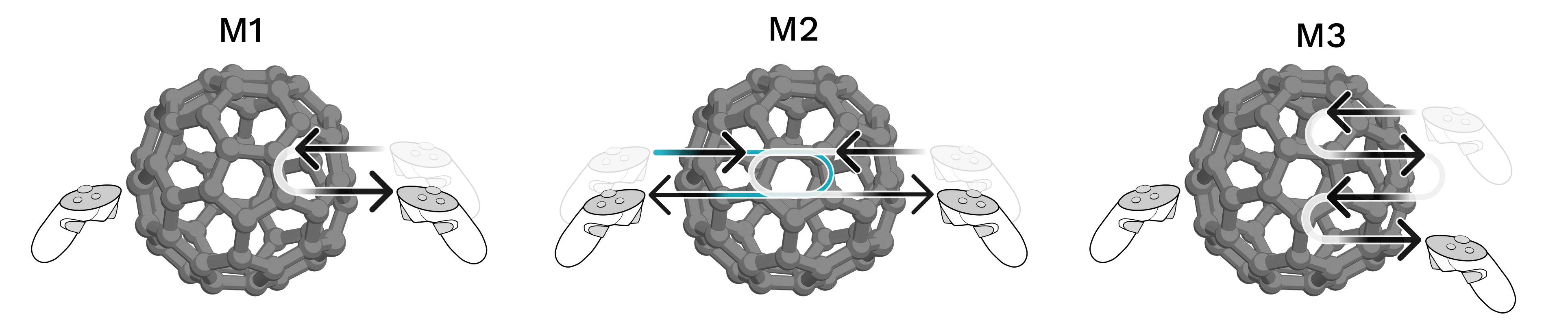}
\vspace{0.1em}
\caption{
     \textbf{Interaction techniques used in the Observing Trials.}
     Illustrations of the three movements used by the experimenter to interact with the C$_{60}$ simulations during the recordings for the Observing Trials in Study One.
     Example videos are provided in the SI.
}
\label{fig:interaction_techniques}
\end{figure}

Recordings were made in a single session by the same experimenter.
Simulations were presented in a randomised order, and the experimenter was blinded to the order of presentation.
% After each trial, the experimenter's response to which buckyball was the softest was recorded.
During each recording, the experimenter applied interactions in a predefined and consistent manner (to within natural human variation) over a 30-second period.
The interaction techniques were selected based on our observations of both novice iMD-XR users during pilot studies and experienced iMD-XR users within our research group. 
The techniques are illustrated in \Cref{fig:interaction_techniques}, and example videos are provided in the SI.
The experimenter performed each movement sequentially on both buckyballs before proceeding to the next movement. 
Each movement began with the experimenter holding the relevant buckyball on both sides, after which they performed one of the following motions:
\begin{enumerate}[label=\textit{M\arabic*:}]
    \item Moving the right hand in and out of the buckyball once.
    \item Moving both hands inwards to cross the arms, then uncrossing them.
    \item Moving the right hand in and out of the buckyball twice.
\end{enumerate}

The left--right positions of the reference and modified buckyballs were counterbalanced, in addition to the buckyball with which the experimenter interacted first, giving four recordings per scaling factor.
% (referred to as \textit{R1}--\textit{R4}).
\nameref{S2_Table} shows the mapping of counterbalanced factors to the name of each recording.
% This resulted in four recordings per scaling factor, and 20 recordings in total: $5$ scaling factors $\times$ $4$ recordings. 
Participants were presented with five recordings per scaling factor, which comprised each of the four recordings and one additional recording selected at random from the set.

% Four recordings were made per scaling factor to counterbalance both the left--right positions of the reference and comparison buckyballs and the buckyball with which the experimenter interacted first. 
% For example, in \textit{R2} for each scaling factor the experimenter interacted with buckyball A first (the reference, softer) and buckyball B was the modified buckyball (more rigid). 

\subsection*{Experimental design}

\subsubsection*{Participants}

\paragraph*{Study One.}
A total of 28 participants were recruited at two Spanish universities: 7 at the Department of Chemistry at the Autonomous University of Madrid and 21 at Centro Singular de Investigación en Tecnoloxías Intelixentes at the University of Santiago de Compostela. 
In Madrid, participants were recruited from the Master’s course in computational chemistry.
In Santiago de Compostela, recruitment was conducted via convenience sampling, primarily through advertisements targeting students and staff in the chemistry and physics departments at the University.
Data were collected between 26 May and 26 June 2025. 
The sample comprised eight women, 19 men, and one non-binary participant, with a mean age of $29 \pm 7$ years. 
Most participants ($n=24$) reported no adverse effects during XR use; three reported headaches, and one reported dizziness. 
One participant had previously taken part in a similar study in our research group.

\paragraph*{Study Two.}
A total of 14 participants were recruited at one Spanish university and one British university: 13 at Centro Singular de Investigación en Tecnoloxías Intelixentes at the University of Santiago de Compostela, and one at the School of Chemistry at the University of Bristol. 
Non-iMD-XR experts (8 participants) were recruited via convenience sampling in the same manner as Study One, and all participated at the University of Santiago de Compostela.
The iMD-XR experts were recruited from our research group (5 participants, including three of the authors of this work) and our network of collaborators (1 participant, Bristol).
Author RRW conducted the experiments and did not participate in the study.
Data were collected between 3 June and 6 August 2025.
The sample comprised five women, seven men, one non-binary participant, and one who preferred not to say, with a mean age of $33 \pm 9$ years. 
Most participants ($n=13$) reported no adverse effects during XR use, and one expert iMD-XR user reported sickness/nausea.
Two non-iMD-XR experts had previously taken part in a similar study in our research group.

\paragraph*{Further information.} Further demographic information is provided in \nameref{S1_Text}.

\subsubsection*{Hardware}

Participants used a Meta Quest 3 VR headset with controllers.
SubtleGame was run using a standalone Android build, except for one participant who used Meta Horizon Link with a laptop running SubtleGame through the Unity Editor.
Each VR headset used a room-scale setup of approximately \SI{2}{\metre} $\times$ \SI{3}{\metre}.
The NanoVer server, game manager and Unity Editor (where applicable) were run on a laptop running Windows 11. 

\subsubsection*{Study procedure}

\begin{figure}
    \centering
    \includegraphics[width=\linewidth]{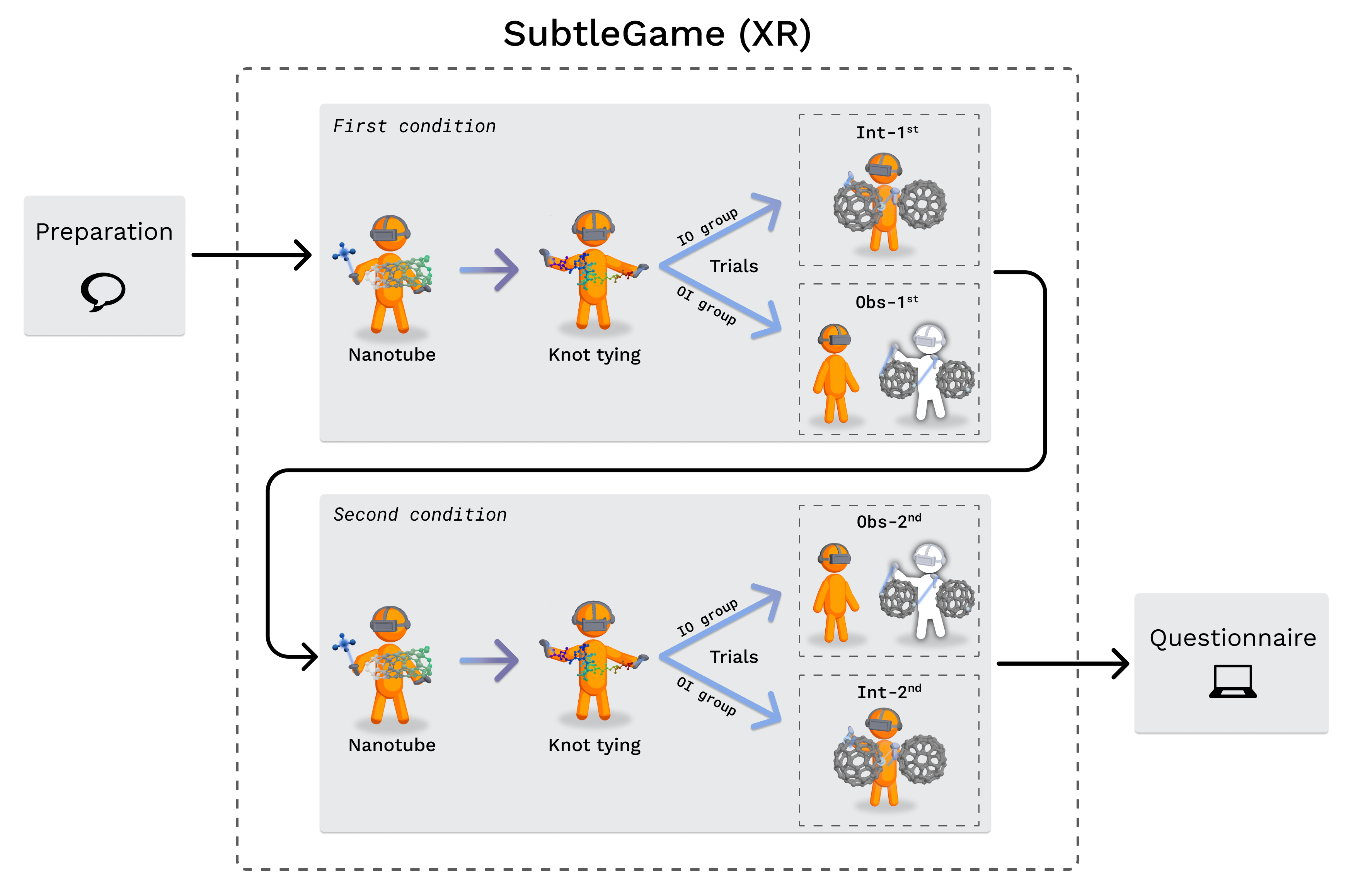}
    \caption{
    \textbf{Study One procedure.}
    The study comprised three sections: the preparation, SubtleGame and the questionnaire.
    % The Interacting and Observing conditions for the Trials were counterbalanced across participants.
    }
    \label{fig:experimental_procedure}
\end{figure}

The procedure of Study One is shown in \Cref{fig:experimental_procedure}.
Study Two followed the same general procedure, except that participants performed the Trials under the Interacting condition in both sections of the game.
Each experiment took approximately one hour and consisted of a preparation phase, SubtleGame (the XR section of the study), and a questionnaire. 
Experiments were conducted in English by the same experimenter, except for one iMD-XR expert in Study Two, who conducted the study independently using written instructions.
The study information sheets from the preparation phase are provided in \nameref{S1_File} and \nameref{S2_File}.

\paragraph*{(1) Preparation.}
The participant read and signed an informed consent form, and were able to ask questions to the experimenter. 
They confirmed that they had normal or corrected-to-normal vision.
They read an information sheet on a computer detailing the SubtleGame tasks.
Following this, the experimenter clarified that, in the Trials task, the softest buckyball is the one that deforms the most easily.
The experimenter also highlighted that, when interacting, the participant needs to deform the buckyballs \textit{inwards} to detect the softness, and that they recommended not only \textit{looking} for the softness but also trying to \textit{feel} it.
The experimenter highlighted the three interaction techniques shown in the videos on the information sheets, recommending that participants try each of the three techniques in the training trials.

\paragraph*{(2) SubtleGame and questionnaire.} 
The participant entered XR and played SubtleGame, starting with a training session in which they learned to interact with an example molecular simulation.
% (with two decane molecules) using the VR controllers. 
The experimenter explained that the participant would not be able to apply a force when their hand is very far from the molecule (as a result of the Gaussian user force).
The participant then played the Nanotube and Knot-tying tasks, followed by the Trials task under their first experimental condition. 
They were offered an optional break before repeating the three tasks in the same sequence, completing the Trials task under their second experimental condition. 
In the training sections of the Trials task involving direct interaction, the experimenter reminded participants to try the three interaction techniques.
After finishing, participants exited XR and completed an online questionnaire.

\subsection*{Data analysis}

This section outlines the analyses conducted for Study One.
The psychometric analysis described here was also applied to Study Two.
Further details are provided in the Supporting Information.

% \subsubsection*{Terminology}

% Each participant was randomly assigned an ID number and, for example, `P1' is used to refer to participant one.
% Trial numbers refer to the position within the full set of 25 trials for the given experimental condition, thus range from 1 to 25 in chronological order as presented in the game. 
% Where relevant, the position within a specified scaling factor is also reported.
% For example, `2 of 5' denotes the second of the five trials for the specified scaling factor for the given experimental condition.

\subsubsection*{Statistical analysis}

Our first aim was to determine whether there were significant differences in performance between the Interacting and Observing conditions in Study One.
Total score distributions during the 2AFC trials were analysed to assess these experimental conditions and learning effects.
% effects of experimental condition (Interacting vs. Observing), overall condition order (first condition vs. second condition), and condition-specific order (Interacting/Observing performed first/second).
Distributions were tested for normality using the D'Agostino \& Pearson’s and Shapiro-Wilk tests, and the results indicated that no distributions differed significantly from normality (\nameref{S4_Table}).
Statistical significance was gauged using \textit{t}-tests, with $p < 0.05$ considered statistically significant.
Paired \textit{t}-tests were used to compare paired distributions, and independent-samples \textit{t}-tests were used for independent distributions.
% Where significant differences were observed, 
Cohen's \textit{d} values indicate the standardised effect size in units of the number of standard deviations.
Average completion times were computed as the mean trial duration per experimental condition and compared using a Wilcoxon signed-rank test.

\subsubsection*{Psychometric analysis}

The just-noticeable difference (JND) for each experimental condition was calculated from the 2AFC results.
JNDs were calculated from the psychometric curves fitted according to the procedure described by Kingdom and Prins \cite{kingdom_chapter_2016}, and give the average threshold at which participants detected rigidity differences of the buckyballs (see \nameref{S7_Text} for further details).
These values were used to quantitatively compare perceptual thresholds between the Interacting and Observing conditions.
% and to understand whether this perception might be useful in detecting rigidity differences in other systems.
% These JNDs indicate the average threshold at which participants detected differences in rigidity between the buckyballs, and were used to quantitatively compare: (a) the impact of the Interacting and Observing conditions on participants' ability to detect these differences (Study One), and (b) differences in perception when comparing hard scaling factors (Study One) and soft scaling factors (Study Two).
% The JNDs were also compared with typical differences between chemical bonds to assess whether these perceptual thresholds fall within a meaningful range for real chemistry contexts.

\subsubsection*{Trajectory analysis}

The iMD-XR trajectories comprise recordings of the MD simulation, including atom positions and user-applied forces at each time point, $t$, during a C$_{60}$ trial.
% which include the positions and velocities of the atoms, the user-applied forces, and other simulation data output by NanoVer, e.g. the potential energy.
Two approaches were employed to examine the trajectories: (a) watching the recordings of the participants performing the Trials task in XR, and (b) plotting the normed user forces and root-mean-square deviation (RMSD).
Watching the recordings facilitated a qualitative, subjective understanding of the interaction techniques used by participants, and the plots enabled quantitative characterisation of these interactions.
The trajectory analysis was performed in Python using MDAnalysis \cite{michaud-agrawal_mdanalysis_2011, gowers_mdanalysis_2016}.

\begin{figure}
    \centering
    \includegraphics[width=\linewidth]{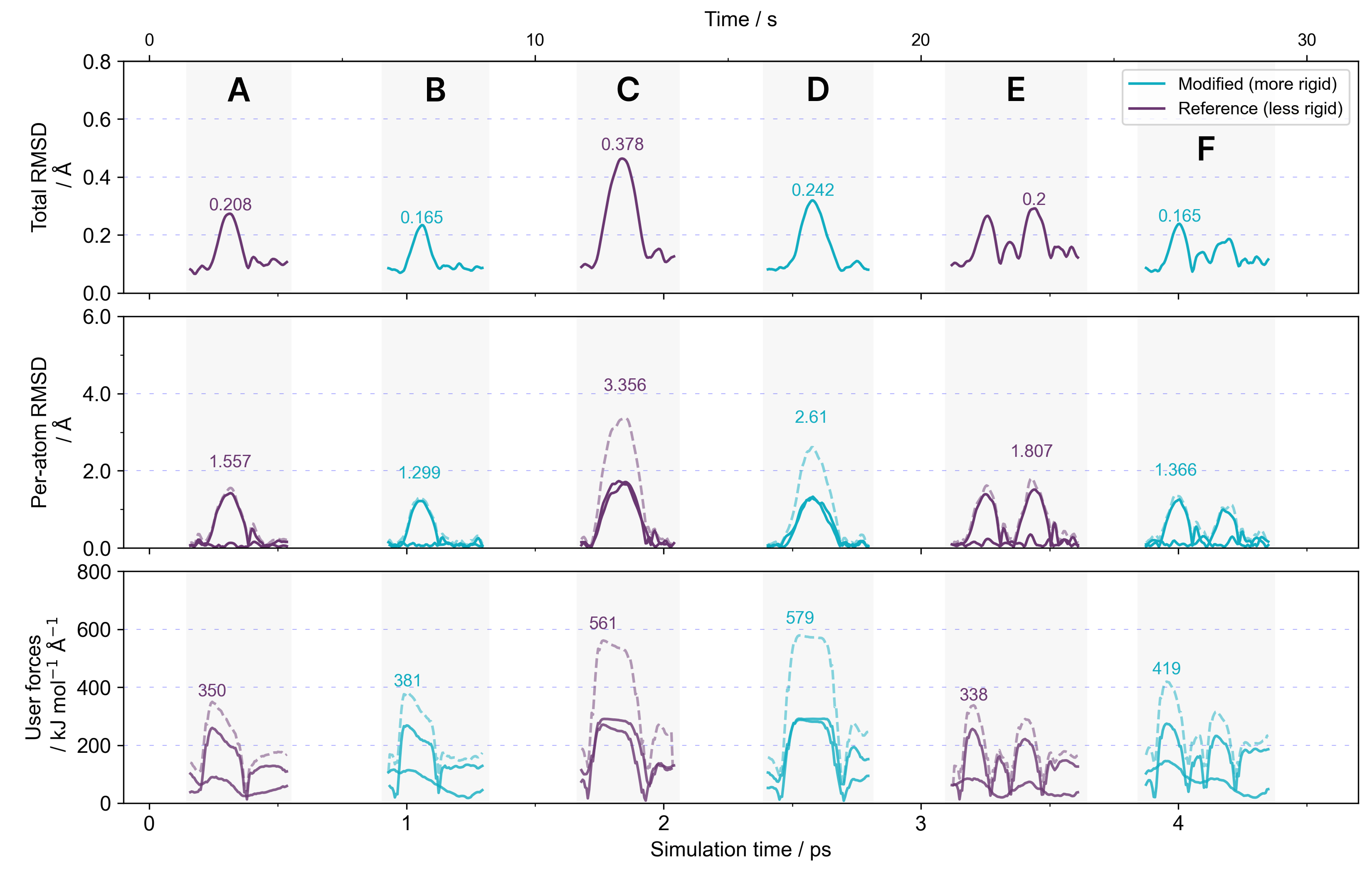}
    \caption{
    \textbf{Total RMSD, per-atom RMSDs of the interacted atoms and normed user forces during Recording 2 for $x=1.375$ in Study One.} 
    Plots correspond to times during which the user is interacting (labelled alphabetically).
    Dashed lined quantities are sums of the solid line quantities beneath them.
    Maximums are labelled for each interaction region, and are relative to the starting value of the given region.
    The bottom axis gives the simulation time in picoseconds and the top axis gives the real-world time in seconds.
    }
    \label{fig:rmsd_example_plot}
\end{figure}

\Cref{fig:rmsd_example_plot} shows an example plot of the total RMSD, per-atom RMSDs and user force magnitudes (Recording 2 for $x=1.375$ of the Observer trials), for which the modified buckyball is more rigid than the reference.
The total RMSD indicates the overall deformation of each buckyball and the per-atom RMSDs indicate the deformations of the atoms the participant interacted with.
The normed user forces indicate the amount of force applied by the participant to the buckyballs.

As described above, the experimenter applies a series of six interactions: \textit{M1} on each buckyball in turn (regions \textbf{A} and \textbf{B}), then repeats with \textit{M2} (regions \textbf{C} and \textbf{D}), then repeats again with \textit{M3} (regions \textbf{E} and \textbf{F}).
Each interaction type is characterised by a specific shape in the user force profile. 
As can be seen in \Cref{fig:rmsd_example_plot}, the total RMSD and per-atom RMSDs are smaller for the more rigid (modified) buckyball than the reference for each interaction type, illustrating how the more rigid buckyball deforms less than the softer one for a given interaction technique.

% and thus is more closely coupled with the pseudo-haptics sense of the interactor. 

\subsubsection*{Questionnaire responses and thematic analysis}
Questionnaires were administered with the Qualtrics platform and contained questions about the Trials task and demographics (\nameref{S3_File} and \nameref{S4_File}).
In Study One, participants were asked to rate the difficulty of the psychophysics trials when interacting and observing on a Likert scale from 1 (\textit{very difficult}) to 5 (\textit{very easy}), and to select which condition they most enjoyed: \textit{Interacting}, \textit{Observing} or \textit{No preference}. 
Participants were then asked to write any further comments that they had about the Trials task.
An inductive thematic analysis was used to categorise the responses into themes.
During preprocessing, grammar and spelling were corrected, empty responses were ignored (ten responses) and two responses were excluded (one due to unclear meaning and one containing only symbols).
The resulting dataset comprised responses from 18 participants (495 words).
% Single sentences that involved multiple themes were assigned to the most relevant theme.
Each response was assigned at least one theme, and all responses are provided in \nameref{S9_Text}.

% --------------------------------------------------- %
\section*{Results}

\subsection*{Total scores}

\begin{figure}
    \centering
    \includegraphics[width=\linewidth]{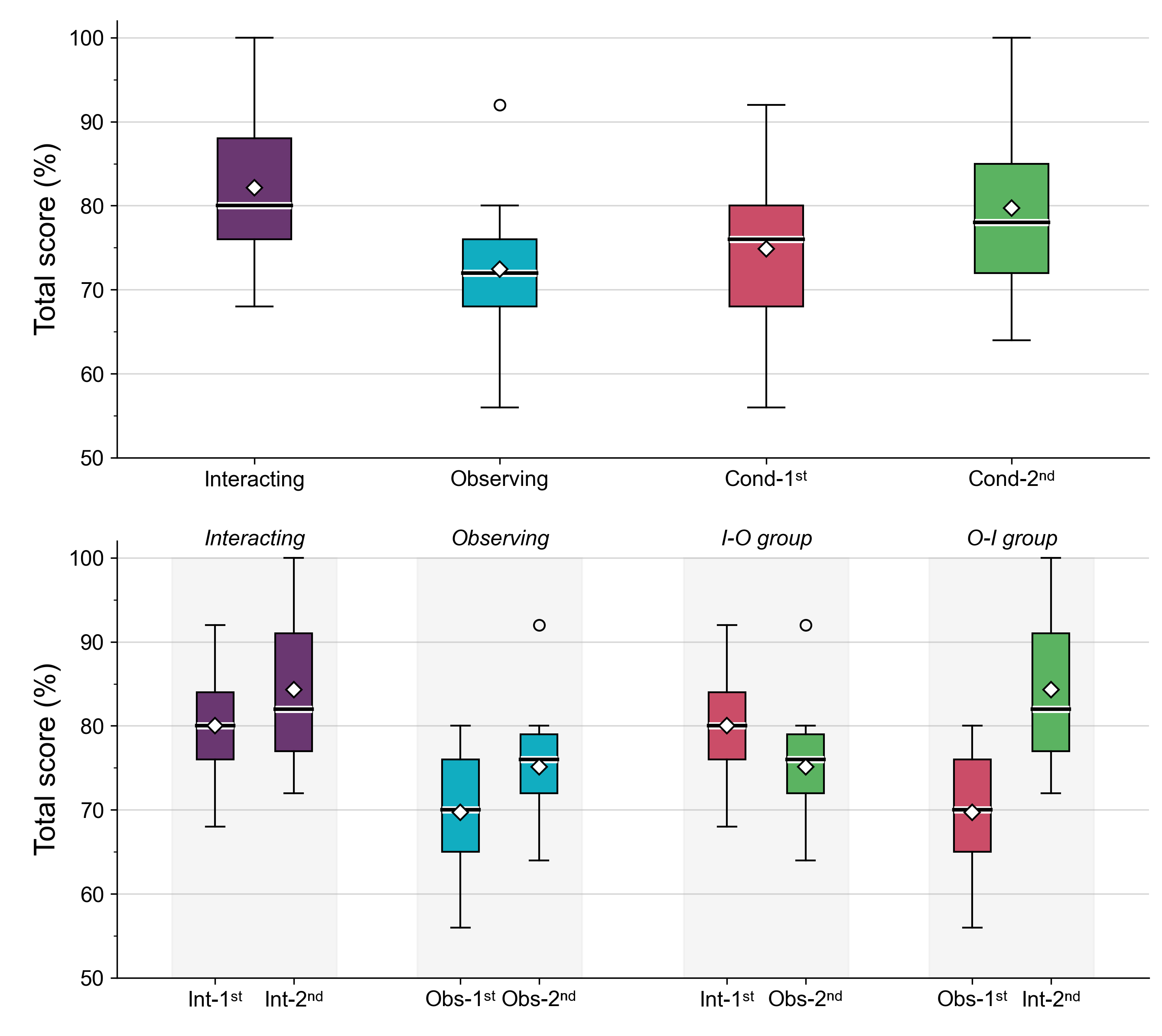}
    \caption{
        \textbf{Total score distributions from the 2AFC Trials for Study One.}
        Distributions are given by experimental condition and first/second condition (top), and by ordered experimental condition and experimental group (bottom).
        The boxes indicate the interquartile range (IQR), the horizontal lines indicate the median, the diamonds indicate the mean, and the circles indicate outliers (values lying more than 1.5 IQR from the nearest quartile).
        The corresponding statistical analysis of these distributions is provided in \Cref{tab:study_one_total_score_significance_testing_parametric}.
    }
\label{fig:box_plots_total_scores}
\end{figure}

% General comparison of Interacting and Observing
The total scores during the 2AFC trials for Study One are reported in \Cref{fig:box_plots_total_scores} and the results of the pairwise comparisons from the $t$-tests are summarised in \Cref{tab:study_one_total_score_significance_testing_parametric} (see the SI for further details).
The results indicate that participants had a better perception of the relative rigidities of the buckyballs for Interacting than for Observing for most comparisons between the two conditions.
Significantly higher scores were seen for Interacting than for Observing when separately comparing the conditions performed first (Int-1\textsuperscript{st} and Obs-1\textsuperscript{st}; $p=0.001$, $d = 1.40$) and the conditions performed second (Int-2\textsuperscript{nd} and Obs-2\textsuperscript{nd}; $p=0.01$, $d=1.11$), both with large effect sizes.
This is also reflected when comparing scores across both the first and second conditions (Int and Obs; $p<0.001$, $d=0.84$), also with a large effect size.
The largest effect size between Interacting and Observing distributions was seen when comparing the first and second conditions within the O-I group (Obs-1\textsuperscript{st} and Int-2\textsuperscript{nd}; $p < 0.001$, $d = 1.46$).
By contrast, the same comparison within the I-O group was the only one for which the Interacting and Observing conditions did not differ significantly.
Although the mean score was higher for Interacting, the $t$-test indicated no significant difference (Int-1\textsuperscript{st} and Obs-2\textsuperscript{nd}; $p=0.13$).
Thus, observers who had already performed the Interacting Trials performed comparably with interactors who had no prior experience of the Trials task.
Collectively, these results suggest that iMD-XR participants can better detect rigidity differences when interacting with molecular simulations than when watching another interact with them.

\begin{table*}
    \centering
    \caption{
        \textbf{Statistical analysis of the total score distributions for Study One.}
        Showing $t$-statistics, $p$-values, and Cohen’s $d$ for the paired and independent sample $t$-tests performed between pairs of distributions shown in \Cref{fig:box_plots_total_scores}.
        Values are given to two decimal places, except where greater precision is needed.
        Statistical significance is indicated by \textsuperscript{*} for $p< 0.05$.
        `Int' and `Obs' denote the Interacting and Observing conditions, respectively, and `1\textsuperscript{st}' and `2\textsuperscript{nd}' indicate whether the condition was performed first or second by participants.
        `Cond-1\textsuperscript{st}' and `Cond-2\textsuperscript{nd}' refer to the first and second experimental conditions performed by participants, irrespective of whether these were Interacting or Observing.
    }
    \renewcommand{\arraystretch}{1.4}
    \setlength{\tabcolsep}{6pt}
    \begin{tabular}{
        >{\centering\arraybackslash}m{1.2cm}
        >{\arraybackslash}m{2.0cm}
        >{\arraybackslash}m{2.0cm}
        S[table-format=2.2]
        S[table-format=<1.3]
        >{\centering\arraybackslash}m{2.0cm}
    }
        \toprule
        \multirow{2}{*}{ }
        & \multicolumn{2}{c}{\raisebox{-2.2ex}{\textbf{Distributions}}}
        & \multicolumn{2}{c}{\textbf{\textit{t}-test}}
        & \textbf{Cohen's \textit{d}} \\
        & & &
        \multicolumn{1}{c}{\textit{t}-stat.} &
        \multicolumn{1}{c}{\textit{p}}      &
        \multicolumn{1}{c}{\textit{d}}      \\
        \midrule
        \multirow{4}{*}{\rotatebox[origin=c]{90}{\textbf{Paired}}}
        & Int    & Obs   &  4.38  & <0.001{\textsuperscript{*}} &  0.84 \\
        & Int-1\textsuperscript{st}   & Obs-2\textsuperscript{nd}  &  1.60  & 0.13                          &  {--}   \\
        & Int-2\textsuperscript{nd}   & Obs-1\textsuperscript{st}  &  5.26  & <0.001{\textsuperscript{*}}   &  1.46 \\
        & Cond-1\textsuperscript{st}  & Cond-2\textsuperscript{nd} & -1.77  & 0.09                          &  {--}   \\
        \midrule
        \multirow{4}{*}{\rotatebox[origin=c]{90}{\textbf{Independent}}}
        & Int-1\textsuperscript{st}   & Obs-1\textsuperscript{st}  &  3.71  & 0.001{\textsuperscript{*}}    &  1.40 \\
        & Int-1\textsuperscript{st}   & Int-2\textsuperscript{nd}  & -1.33  & 0.20                          &  {--}   \\
        & Int-2\textsuperscript{nd}   & Obs-2\textsuperscript{nd}  &  2.95  & 0.01{\textsuperscript{*}}     &  1.11 \\
        & Obs-1\textsuperscript{st}   & Obs-2\textsuperscript{nd}  & -2.08  & 0.048{\textsuperscript{*}}    &  0.78 \\
        \bottomrule
    \end{tabular}
    \vspace{0.9em}
\label{tab:study_one_total_score_significance_testing_parametric}
\end{table*}

% Learning effects
We investigated learning effects to examine whether performance improved across the game.
On average, participants scored higher in their second condition, however, this difference was not statistically significant (Cond-1\textsuperscript{st} and Cond-2\textsuperscript{nd}; $p = 0.09$).
% Given that the \textit{p}-value is close to the significance threshold, this result suggests that there may be a learning effect in which performance improves over the course of the game, however, more data would be needed to confirm this.

We also investigated training effects to examine how each condition influenced performance in the other condition. 
To do this we compared performance in the same condition (Interacting or Observing) between the two groups (I-O and O-I).
The results of comparing the two Observing distributions suggest that the Interacting condition trained participants to perform better in the Observing condition: Obs-2\textsuperscript{nd} observers performed significantly better than Obs-1\textsuperscript{st} observers (Obs-1\textsuperscript{st} and Obs-2\textsuperscript{nd}; $p = 0.048$, $d = 0.78$).
In contrast, there was no statistically significant difference between the Int-1\textsuperscript{st} and Int-2\textsuperscript{nd} interactors (Int-1\textsuperscript{st} and Int-2\textsuperscript{nd}; $p = 0.20$).
% However, Int-2\textsuperscript{nd} interactors performed better on average than the Int-1\textsuperscript{st} interactors, suggesting that there may be an effect, which would require further sampling to investigate.
% Thus, whilst observing does not appear to train us to better detect rigidity differences when interacting, interacting appears to train us to observe these differences more effectively.

\Cref{fig:box_plots_total_scores} shows one participant (P25) who achieved a markedly higher score in the Observing condition (\SI{92}{\percent}) than the rest of the cohort, which is commented on in \nameref{S6_Text}. 

% This participant scored lower in the Interaction condition (\SI{76}{\percent}), which they performed first.
% This data point may reflect individual differences in perceptual sensitivity, however, no firm conclusions can be drawn.
% Further studies with larger sample sizes would be required to determine whether a subgroup of participants systematically performs better under Observing conditions than the Interacting condition.
% % Although this performance exceeds the typical range observed for Observing, it does not alter the overall conclusions from the statistical analysis (the outlier was included in the analysis).

\subsection*{Participant interactions during the Interacting trials}

We investigated participants' interactions with the buckyballs during the Interacting trials of Study One to examine: (a) whether the manner in which participants interacted influenced their score, and (b) whether participants’ interactions were comparable to those of the experimenter during the recordings used in the Observing trials.
Since the dataset comprised a large number of trajectories (25 trials $\times$ 28 participants $=$ 700 trajectories), our analysis focused on the participants who scored the highest (P1 and P11, both scoring $100 \%$) and the lowest (P9 and P24, both scoring $68\%$) in the Interacting condition.

\begin{figure}
    \centering
    \includegraphics[width=0.98\linewidth]{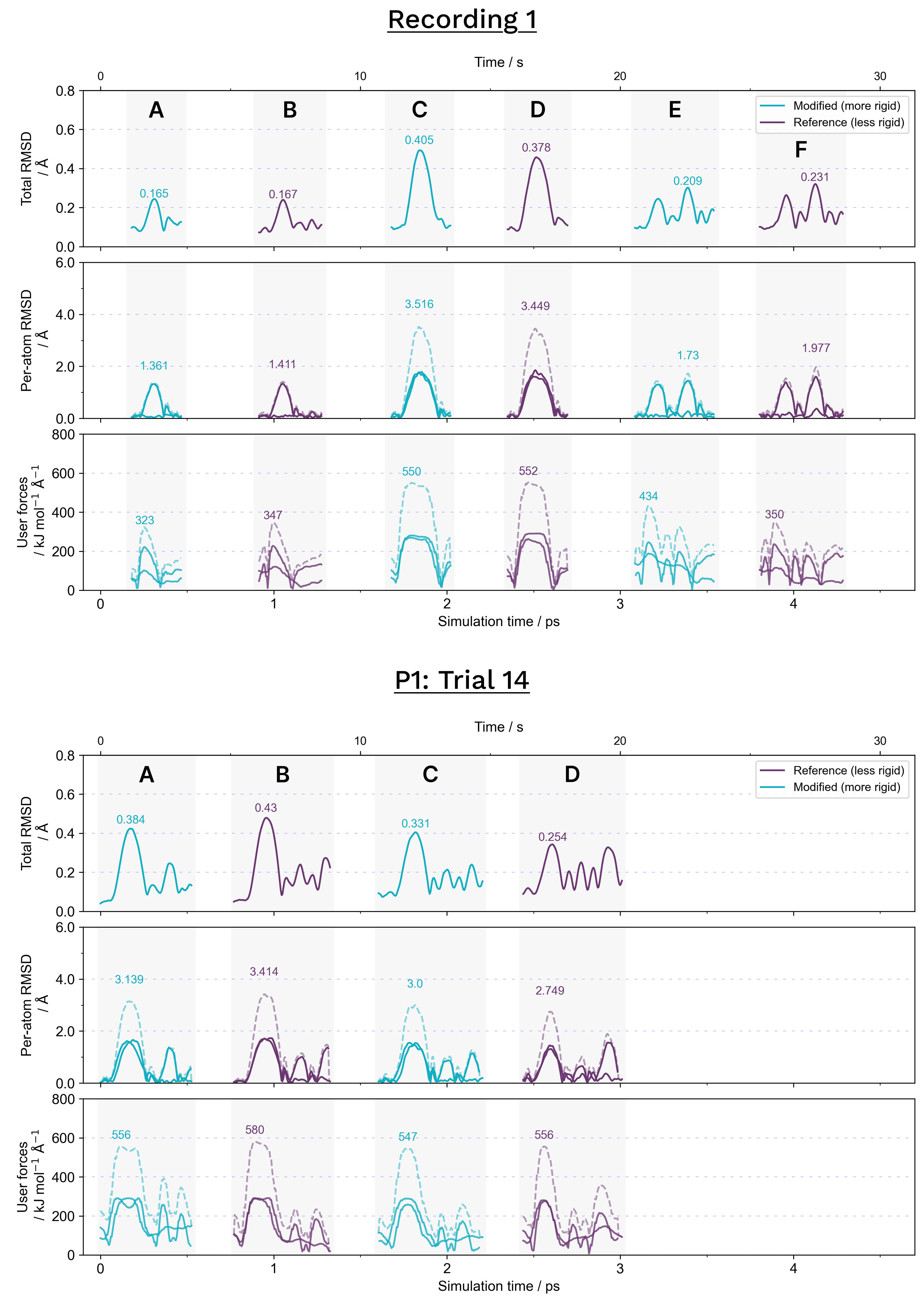}
    \vspace{0.1em}
    \caption{
        \textbf{Total RMSD, per-atom RMSDs of the atoms the user is interacting with, and user force magnitudes for $x=1.036$ in Study One during Recording 1 (top) and trial 14 for P1 (bottom).}
        Plots correspond to times during which the user is interacting (labelled alphabetically).
        Dashed lined quantities are sums of the solid line quantities beneath them.
        Maximums are labelled for each interaction region, and are relative to the starting value of the given region.
        The bottom axis gives the simulation time in picoseconds and the top axis gives the real-world time in seconds.
    }
\label{fig:rmsds_and_user_forces_P1_and_recordings}
\end{figure}

\begin{figure}
\centering
    \includegraphics[width=0.98\linewidth]{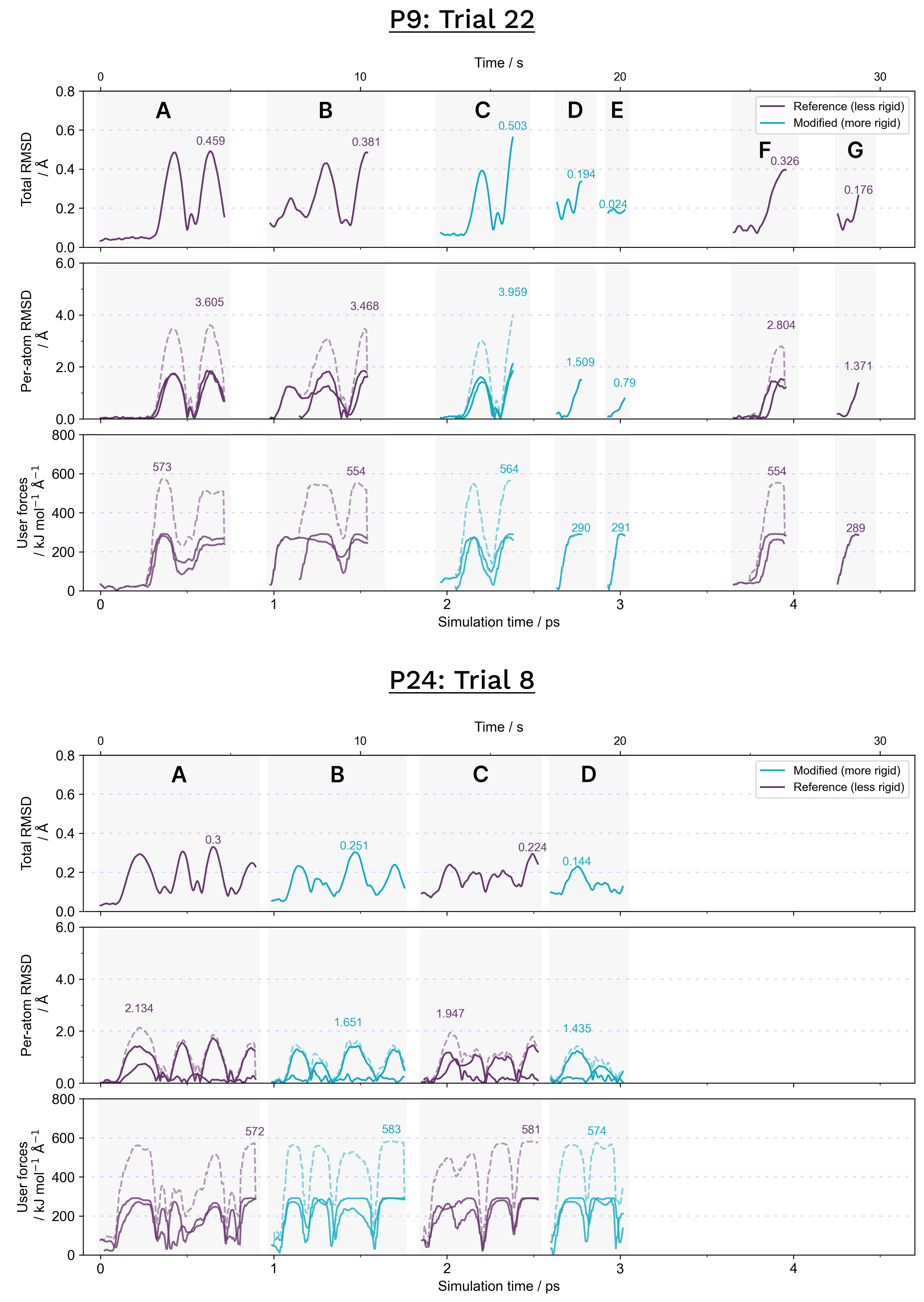}
    \vspace{0.1em}
    \caption{
        \textbf{Total RMSD, per-atom RMSDs of the atoms the user is interacting with, and user force magnitudes for $x=1.036$ in Study One during trial 22 for P9 (top) and trial 8 for P24 (bottom).}
        Plots correspond to times during which the user is interacting (labelled alphabetically).
        Dashed lined quantities are sums of the solid line quantities beneath them.
        Maximums are labelled for each interaction region, and are relative to the starting value of the given region.
        The bottom axis gives the simulation time in picoseconds and the top axis gives the real-world time in seconds.
    }
\label{fig:rmsds_and_user_forces_P9_and_P24}
\end{figure}

For the Interacting condition, the score differences between the highest and lowest scorers were largest for the most difficult trials ($x=1.036$), therefore we investigated trajectories for this scaling factor.
\Cref{fig:rmsds_and_user_forces_P1_and_recordings} and \Cref{fig:rmsds_and_user_forces_P9_and_P24} show RMSD and user force profiles for a recording (Recording 1, where the experimenter interacts with the modified ball first), and one representative trial each from P1, P9 and P24 for $x=1.036$.
% Representative plots of the RMSDs and user forces are provided in the SI.

% This indicates the importance of the user forces.
% The magnitude of the user force contributes to the maximum deformation: the magnitude needs to be large enough to cause the buckyball to deform. 
% However, if the force magnitude is large but in a direction that the buckyball cannot move due to its constrained structure, then the deformation will not be as large as one might expect from information about only the force magnitude.
% Furthermore, inconsistently applied forces may result in less clear trends in the deformation. 
% These observations suggest that ability to discriminate small rigidity differences is not based purely on the visual deformation.

\subsubsection*{Highest scorers}

We observed two differences between the highest and lowest scorers: the interaction techniques that they employed and the consistency with which they applied them.
In general, the highest scorers---P1 (\Cref{fig:rmsds_and_user_forces_P1_and_recordings}) and P11 (\nameref{S2_Figure})---interacted with the molecules in a manner similar to the recordings.
Both participants used all three recommended interaction techniques across their Trials, with \textit{M2} being the most frequently used.
Moreover, the maximum user force applied by these participants was generally comparable to those in the recordings when matched by interaction technique, indicating that these participants learned to effectively apply forces to the molecules.
    % \item We noticed that Technique Two tends to be the easiest to perform of the three (participants seem to have more control over the molecules when interacting in this way), which is perhaps why it was favoured by these participants.
These participants also tended to alternate between the two buckyballs in a systematic way: interacting with one buckyball, then repeating the same interaction on the other.
This sequence mirrored the approach used during the recordings, which was chosen because (a) repeating the same interaction on each buckyball facilitates direct comparison, and (b) switching between molecules provides time for each buckyball to relax toward its equilibrium geometry, thus reducing visual noise that may interfere with the task.

\subsubsection*{Lowest scorers}

\Cref{fig:rmsds_and_user_forces_P9_and_P24} provides representative figures of the RMSD and user force profiles for the lowest scorers.
Although P9's maximum user forces and RMSDs were similar to those of the recordings and highest scorers, they lacked consistency in their interactions.
This is most visible when watching from inside VR, however, it is also indicated by the RMSDs and user force profiles.
% Where the plots in \Cref{fig:rmsds_and_user_forces_P1_and_recordings} show consistent interactions applied to the each buckyball in turn, P9 interacted twice with the same buckyball (regions \textbf{A} and \textbf{B}), before moving onto the second buckyball (\textbf{C}--\textbf{E}), then moving back to the first buckyball (regions \textbf{F} and \textbf{G}). 
Although some of the user force profiles exhibit similar shapes, they are not applied to each buckyball in a consistent manner: P9 interacted twice with the same buckyball (regions \textbf{A} and \textbf{B}), before moving onto the second buckyball (\textbf{C}--\textbf{E}), then moving back to the first buckyball (regions \textbf{F} and \textbf{G}), mixing different interaction techniques across the interaction regions. 
% which show how P9 applied two interactions in a row to the same buckyball (\textbf{A} and \textbf{B}), 
% varied shapes of the user force profiles across the interacting regions, e.g. regions \textbf{A} and \textbf{B} show very differen. 
The corresponding RMSDs also differed in form across interactions, illustrating how the resulting deformations may have been difficult to interpret.

Although P24's interactions showed some consistency, they caused only small deformations in the buckyballs for the first two trials for $x=1.036$, suggesting that they may have still been learning how to interact with the molecules during earlier trials.
Shown in \Cref{fig:rmsds_and_user_forces_P9_and_P24}, which was P24's third trial for $x=1.036$, P24 then began to apply large amounts of force to the buckyballs, as shown by the long interaction regions containing a large number of high-force peaks.
This approach may have added visual noise by causing the buckyballs to vibrate and move in chaotic ways, possibly making the task more difficult.
Furthermore, this participant ended all five trials for $x=1.036$ earlier than the maximum time limit.
% For P24, the maximum force in each interaction region is similar to the highest scorers and recordings, however, the resulting deformations are low. 
% This suggests that this participant was not able to apply forces to the buckyballs in a way to cause deformations, indicating that they were still been learning how to interact with the molecules.

Both participants also used self-devised interaction techniques, rather than choosing from the recommended ones.
In particular, P9's self-devised technique involved attempting to hold the buckyball in the maximally deformed position, which often caused the buckyballs to spin, likely making the task more difficult.
Furthermore, this choice of technique suggests that P9 was focusing on the static deformation (i.e., the instantaneous distance between the two interacted atoms) rather than the \textit{dynamic} response of the system.
This may have hindered performance, since the task was to compare the rigidity of the molecules---an inherently dynamic property.

\subsubsection*{General observations}

The trajectories show that both the highest and lowest scorers were able to apply close to the theoretical maximum user force magnitude calculated in \nameref{S4_Text}. 
However, it is clear from the RMSDs that applying large forces did not always induce large deformations in the buckyballs (e.g. P24 in \Cref{fig:rmsds_and_user_forces_P9_and_P24}), further indicating the importance of interaction technique in the 2AFC task.
Furthermore, the maximum total RMSD values seen in both the participants' trajectories and the Observing condition recordings were consistently lower than the average maximum deformation calculated in \nameref{S4_Text}.
This is unsurprising, given that participants used a force that varied with distance, which was applied using XR controllers, whereas the automated protocol employed a constant (maximum) force from a consistent position to achieve maximal buckyball deformation.

\subsection*{Psychometric analysis}

\begin{figure}
\centering
    \includegraphics[width=\linewidth]{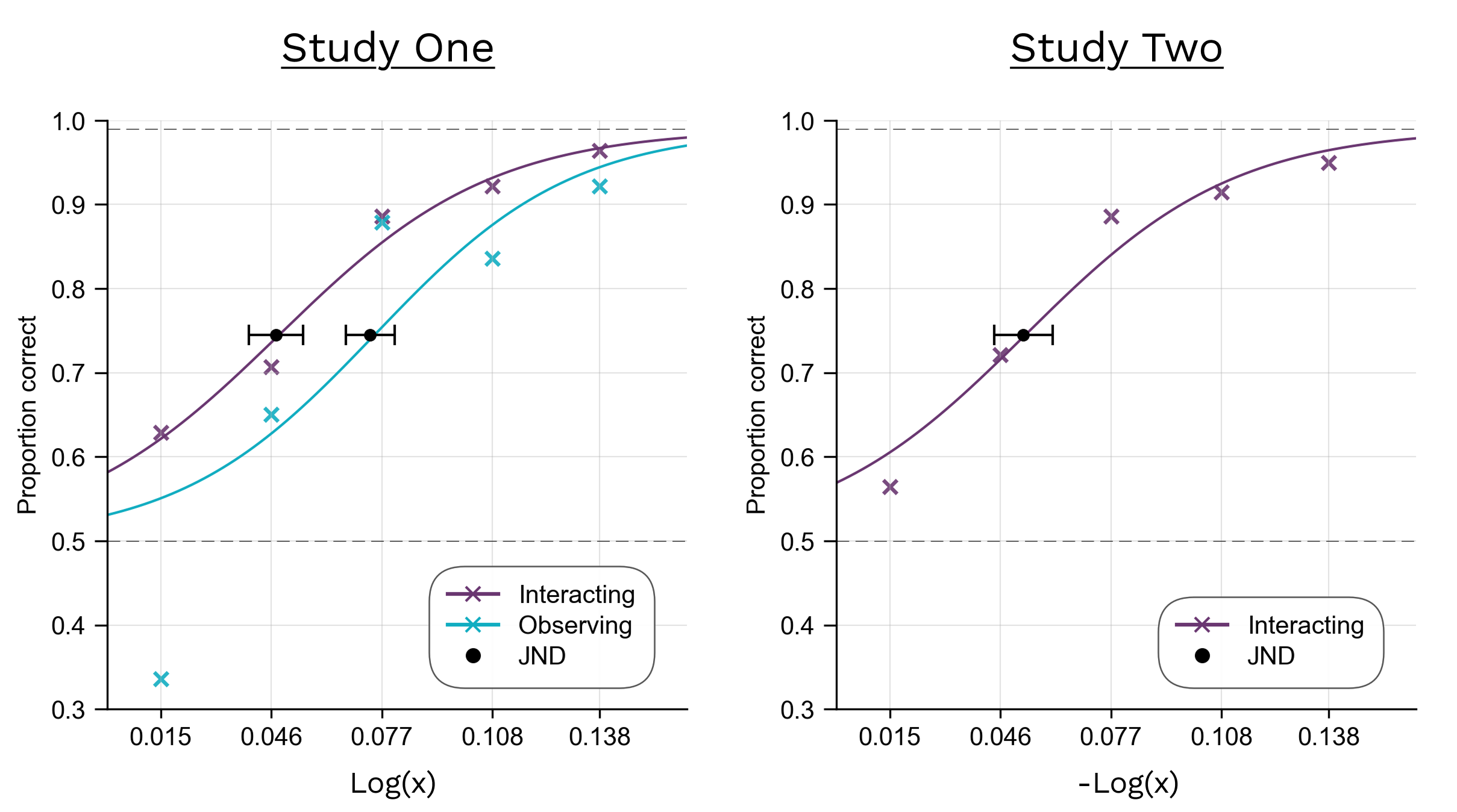}
    \vspace{0.1em}
    \caption{
        \textbf{Psychometric curves for Study One and Study Two.}
        Variable $x$ is the rigidity scaling factor.
        The scatter plot data show the average probability across the participants for each scaling factor and the curves were fitted to these data using a Bayesian fitting procedure for illustrative purposes only.
        JND values are given with one standard deviation error bars.
    }
\label{fig:psychometric_curves}
\end{figure}

\Cref{tab:jnd_results} reports the JNDs for Studies One and Two.
To enable comparison between the Interacting and Observing conditions, the thresholds are expressed as Weber fractions, which quantify the relative change in rigidity required for successful discrimination and are defined as
\begin{equation}
    K = \frac{\Delta k}{k},
    \label{eq:weber_fraction_k}
\end{equation}
where $k$ is the reference angle force constant and $\Delta k$ is the absolute difference at the threshold. 
As Weber fractions, the JNDs were $0.115$ (Study One) and $0.114$ (Study Two) for the Interacting condition, and $0.185$ (Study One) for Observing.
The JNDs are also shown in \Cref{fig:psychometric_curves}, along with the psychometric curves fitted to the average probabilities across the participants.

\begin{table}
    \centering
    \caption{
    \textbf{JNDs for Studies One and Two.}
    JNDs are reported as (a) scaling factors of the angle force constant, $x$, with upper and lower bounds, (b) Weber fractions, $K$, and (c) differences in angle force constant, $\Delta k$. 
    Values of $x$ and $K$ are given to three decimal places, and $\Delta k$ to the nearest whole number. 
    The reference angle force constant was $k = \SI{457.672}{\kilo\joule\per\mol\per\radian\squared}$.
    Upper and lower bounds were calculated from the standard deviation of the fitted curves, which were converted from $\log$ space.
    `Int.' refers to the Interacting condition and 'Obs.' to the Observing condition.
    }
    \renewcommand{\arraystretch}{1.5}
    \setlength{\tabcolsep}{6pt}
    
    \begin{tabular}{
        >{\centering\arraybackslash}m{1.2cm}
        >{\centering\arraybackslash}m{2.0cm}
        >{\centering\arraybackslash}m{1.3cm}
        >{\centering\arraybackslash}m{2.5cm}
        >{\centering\arraybackslash}m{1.0cm}
        >{\centering\arraybackslash}m{2.6cm}
    }
    \toprule

     &
    \multirow{2}{*}{\makecell[c]{\textbf{Scaling}\\\textbf{factors}}} & \multirow{2}{*}{\textbf{Cond.}} & $x$ & $K$ & $\Delta k$ \\
    
     & & & (range) & & \si{\kilo\joule\per\mol\per\radian\squared} \\
    
    \midrule
    
     \multirow{2}{*}{\makecell{\textbf{Study}\\\textbf{One}}}
      & \multirow{2}{*}{\makecell{More rigid\\($>1$)}} & Int. & 1.115 (1.096--1.135) & 0.115 & 53 \\
      & & Obs. & 1.185 (1.166--1.204) & 0.185 & 85 \\
    
    \midrule
    
     \textbf{Study Two}
     & Less rigid ($<1$) & Int. & 0.886 (0.870--0.903) & 0.114 & $-52$ \\
    
    \bottomrule
    \end{tabular}
    \label{tab:jnd_results}
\end{table}

Of note is the unusually low probability for the smallest scaling factor under the Observing condition ($p = 0.33$), which is commented on in \nameref{S8_Text}.

\subsection*{Enjoyment and difficulty ratings}

\begin{figure}[htbp]
    \centering
    \includegraphics[width=\linewidth]{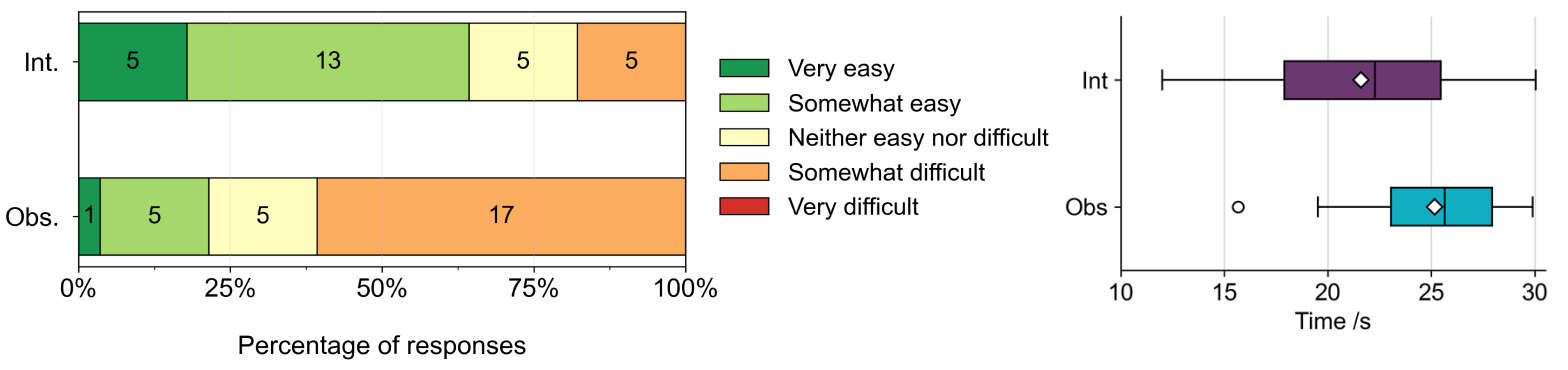}%
    \caption{
        \textbf{Difficulty ratings (left) and average time taken per trial (right) for Study One.}
    }
    \label{fig:difficulty_ratings}
\end{figure}

Participants generally preferred the Interacting condition, with 27 participants (\SI{96}{\percent}) reporting that they enjoyed interacting more than observing, and one having no preference. 
% P25 reported that they had no preference between the conditions. 
% This is the same participant mentioned previously, who achieved a markedly higher score in the Observing condition.
The Observing condition was rated as more difficult overall: the majority of participants rated Observing as \textit{somewhat difficult} ($n = 17$), whereas Interacting was most frequently rated as \textit{somewhat easy} ($n = 13$), as shown in \Cref{fig:difficulty_ratings}.
A similar pattern was seen in the average completion time per trial, where participants took longer on average to complete trials in the Observing condition ($25.2 \pm 4.8$~\si{\second}) than in the Interacting condition ($21.6 \pm 5.6$~\si{\second}).

% I don't think I can use a Wilcoxon test here, because the data are cut off at 30s, aren't normal.
% This difference was statistically significant ($z=41$, $p\ll0.001$).

\subsection*{Thematic analysis}
We performed a thematic analysis on the free-form text responses in the questionnaire (\nameref{S9_Text}).
The themes and associated number of statements that emerged from this analysis were: Positive Feedback (5 responses), Interacting Easier or Observing More Difficult (4 responses), Visuo-haptic Illusions (4 responses), Technical Difficulties (4 responses), Fatigue (2 responses) and Interaction Technique (2 responses).
This analysis aided the interpretation of the other results, and participant responses are interwoven into the following discussion.

% --------------------------------------------------- %
\section*{Discussion}
\label{sec:discussion}

% Our results provide quantitative evidence that iMD-XR participants can perceive differences in the mechanical properties of molecular simulations.
% In this section we first examine differences in performance between the Interacting and Observing experimental conditions, showing that interaction facilitated better discrimination performance and improved performance during subsequent observation. 
% Following this we highlight reports by participants of visuo-haptic sensations and indicate future research avenues for investigating these sensations.
% Next, we consider other factors that may have influenced performance, including interaction technique, understanding of the interaction potential and fatigue.
% We then compare the implications of perceiving the properties of virtual objects in iMD-XR with pseudo-haptic approaches involving visuomotor incongruence.
% Following this, we highlight the potential of iMD-XR in chemistry research and education, and discuss how these findings may inform the design of other XR systems.
% Finally, we consider the limitations of this work and propose improvements to the study design and future research directions. 

% Our findings show that XR participants can sense rigidity differences of purely virtual objects without physical feedback, and that this ability is enhanced when interacting compared to observing the interactions of another participant. 

\subsection*{Interaction enabled perception of more subtle rigidity differences}
% \subsection*{Better performance when interacting than observing}

Overall, interactors scored significantly higher than observers.
In addition, of the four comparisons between subsets of the Interacting and Observing distributions, three showed significantly higher scores for Interacting, while one showed no significant difference, although the mean Interacting score was higher.
The JNDs calculated from the psychometric curves were also lower for Interacting than Observing, indicating that interactors could perceive more subtle rigidity differences than observers.

% We hypothesise that the ability to integrate proprioceptive information with their vision
% Multisensory training is known to lead to improved learning compared with unisensory training \cite{shams_influences_2011, shams_benefits_2008, seitz_sound_2006}, which may explain the better performance seen when interacting.
% % tthat interacting provides a richer multisensory experience than observation alone.
% During interaction, participants can integrate proprioceptive information with their vision to understand the molecular objects, whereas observers only have access to the visual information.
% This aligns with findings from neuroscience, where multisensory training is known to lead to improved learning compared with unisensory training \cite{shams_influences_2011, shams_benefits_2008, seitz_sound_2006}. 

In the questionnaire feedback, most participants reported finding it easier to discriminate rigidity differences when interacting than when observing.
Additionally, on average participants spent less time in the Interacting trials than the Observing trials.
% One factor that may have contributed to this difference is the rigidly timed structure of the recordings.
It is important to note that during Observing trials participants needed to wait for the experimenter's avatar to complete the interactions at the pre-set pace of the recordings, whereas in the Interacting condition they could proceed at their own pace, which may have contributed to the increased time required to complete the trials.
Nevertheless, the results illustrate that higher scores were obtained in less time when interacting than observing.

% Our comparison of the JNDs to differences in the bond angle force constants defined in the 17-alanine simulation suggests that XR participants would be able to discriminate C$_{60}$ structures whose rigidities are defined by the 17-alanine angle force constants. 
% Furthermore, the results suggest that interaction would enable participants to discriminate between more pairs of force constants than observation alone.
% It is uncertain whether these findings extend to systems other than the C$_{60}$ structure.
% Detecting rigidities within more flexible structures consisting of multiple atom types and force constant values may prove more challenging.
% % which is highly symmetric and rigid, and comprises a single atom type and a single unique angle force constant.
% % As such, further studies are required to test these comparisons and to quantify perception within other molecular systems.

Further work is required to understand the mechanisms underlying the performance advantage observed during direct interaction.
One possibility is participants were able to integrate proprioceptive information with the visual feedback during interaction, thus providing additional information for interpreting molecular motion.
Another contributing factor may have been attention: participants generally found interaction more enjoyable, and therefore may have been more attentive to the task and consequently performed better.
Somewhat surprisingly, participants were able to discriminate differences in cases where the RMSDs did not display the expected trend, i.e. where the more rigid buckyball deformed less than the other according to the RMSD profiles. 
This may be because the RMSDs provide a compressed representation of molecular deformation, reducing complex three-dimensional structural changes to a single scalar quantity.
This could also suggest that rigidity discrimination is not based solely on the visual deformation of the molecules.
Perhaps participants were sensitive to other aspects such as the exact motion of the XR controller during force application, or, in the case of direct interaction, proprioceptive cues.

\subsection*{Interacting improved subsequent performance when observing}

Multisensory training is known to enhance learning compared with unisensory training, even for tasks that ultimately rely on a single sensory modality \cite{shams_influences_2011, shams_benefits_2008, seitz_sound_2006}. 
Consistent with this, observers performed significantly better after completing the Interacting trials than when observing without prior experience, suggesting that interaction trained participants to observe rigidity differences more effectively. 
% In contrast, performing the Observing trials first did not significantly influence performance in the Interacting trials.
An open question for the application of iMD-XR in chemistry is whether this effect extends beyond iMD-XR workflows. 
If interaction improves the perception of molecular properties during observation, it may also enhance the interpretation of the standard, non-interactive molecular trajectories used in computational chemistry.

% For example, Seitz et al.~\cite{seitz_sound_2006} reported improved performance in a visual task following audiovisual training compared with visual-only training. 
% This aligns with our findings, where multisensory interaction enhanced subsequent performance in the predominantly unisensory Observing task.

\subsection*{Visuo-haptic illusions during interaction}

% In addition to quantitative differences in performance,
Several participants reported `feeling' the virtual objects when interacting with them.
In their questionnaire responses, three participants mentioned `feeling' the softness of the buckyballs when interacting, e.g. ``\textit{a `resistance feeling' even though there is nothing physical to actually feel}'' (P5).
These responses highlight how, alongside being able to perceive differences in molecular properties, participants may also experience visuo-haptic illusions in which they feel haptic-like sensations.

% Although pseudo-haptic sensations are described as illusions, a recent study involving EEG measurements showed that virtual touch in VR---where participants were touched by a virtual robot arm without physical feedback---was processed similarly to physical touch, activating sensorimotor regions in the brain characteristic of physical touch despite the absence of direct physical stimulation \cite{savalle_does_2026}.
% This is an especially intriguing finding in the context of molecular simulations, given that there is no physical analogue for touching molecules. 
% Does the same kind of brain occur when interacting with molecules in iMD-XR, given that we have no prior experience of interacting with them?
% % If this kind of brain activation is seen in participants during an iMD-XR session, then this would suggest that iMD-XR facilitates an experience otherwise not possible in the physical world: \textit{touching} molecules.

% This is interesting in the context of molecular simulations, given that we have no experience of touching molecules in the physical world.
% As a result of this, we have weak priors when it comes to our expectations of how molecules should behave.
% This brings up a question of what brain activation occurs when we interact with molecular simulations given that we have no physical analogue for this experience.

In contrast to the reports of visuo-haptic illusions from other participants, P12 stated that it was ``\textit{a bit tricky to feel the softness of a molecule}'', suggesting that this participant may not have experienced these illusions.
% This observation is in line with the pseudo-haptics literature, 
Subjective experiences of pseudo-haptic sensations are known to be influenced by factors such as prior experience and biological differences and thus vary across individuals \cite{collins_pseudo-haptics_2019}.
% \item However, since molecular properties do not depend upon visuo-haptic illusions in iMD-XR, there is a question of whether the participant's \textit{perception} of molecular properties remains dependent on these illusions.
Interestingly, although they reported difficulty `feeling' the softness, P12 scored higher when interacting (\SI{92}{\percent}) than observing (\SI{72}{\percent}).
This implies that visuo-haptic illusions of `stiffness' or `rigidity' may not be necessary for perceiving rigidity differences.
Moreover, this suggests that interacting may remain more effective than observing for discriminating rigidities even for individuals who do not report experiencing such sensations; however, further investigation is required to test this hypothesis.
% which could include further analysis of the phenomenological experiences of individuals during iMD-XR.

\subsection*{Factors affecting performance}

\subsubsection*{Interaction technique}

To better understand the finding that interacting led to higher performance than observing, we analysed how participants interacted with the buckyballs during the 2AFC trials.
Analysis of the trajectories of the highest- and lowest-scoring participants in the Interacting condition indicated that performance was influenced by both the choice of interaction technique and the consistency with which interactions were applied.
The highest scorers interacted in a similar manner to the experimenter in the Observing trials: they employed the recommended techniques, applied forces consistently, and systematically alternated between the two buckyballs.
In contrast, the lowest scorers deviated from this pattern.
Both adopted self-devised interaction strategies rather than the recommended techniques, with one participant gradually refining their technique across the trials and the other focusing on the static deformation rather than the dynamic response of the buckyballs.

These findings suggest that differences in performance within the Interacting condition were not determined solely by perceptual sensitivity, but were also influenced by \textit{how} participants interacted with the molecules.
This indicates that the Interacting JNDs calculated from Study One may partially reflect variability in interaction skill, and not purely the perceptual thresholds.
%particularly for participants who completed the Interacting trials first.

Poor interaction technique may have been a consequence of poor understanding of the interaction force.
From our observations, it was clear that many participants struggled to fully understand the force that they were applying to the molecules. 
We observed that many participants attempted to interact with the buckyballs at large distances (where the force falls to zero), indicating a lack of understanding about the force that they were applying.
This issue was highlighted by P20, who commented in their questionnaire responses that they were unsure if they were “\textit{properly aware of the strength that I was exerting over the molecule}” and recommended incorporating an exercise to understand this better.

In Study Two, several experienced iMD-XR participants were included and the number of trials was increased to reduce the impact of poor interaction technique on the sampling.
Since Study Two sampled decreasing rigidities, the measured JND was inverted to obtain the equivalent threshold for increasing rigidities (see \nameref{S4_Equation}), enabling direct comparison with the Study One JND. 
This yielded a predicted threshold of \SI{12.9}{\percent}, which is slightly higher than the Interacting threshold from Study One (\SI{11.5}{\percent}), but remains lower than the Observing threshold (\SI{18.5}{\percent}). 
These findings suggest that Study One provides a reasonable approximation of perceptual sensitivity in the Interacting condition, although the true threshold may be slightly higher.

Our analysis also revealed that score differences between the highest and lowest scorers were largest for the most difficult trials and smaller for the easier trials.
This suggests that interaction technique becomes especially important when rigidity differences are small.
This is likely because large rigidity differences are visually obvious, meaning that any interaction that perturbs the buckyballs is sufficient for comparison.

\subsubsection*{Fatigue and attention}

In addition to interaction-related factors, two questionnaire responses indicated that fatigue effects may also have influenced performance, which may be reduced through the addition of more breaks. 
% Two participants reported feeling progressively tired over the course of the game, which may have negatively affected their performance.
% This fatigue could be due to a combination of factors, including wearing the VR headset and being in VR, or from the monotony of the psychophysics task. 
% Although participants were offered a break halfway through the game, fatigue from use of the VR headset may be reduced through the addition of more breaks.
% Participants were offered a break halfway through the game, however, additional breaks may help maintain participant focus in future studies.
One participant also reported gradually losing interest throughout the study.
Since the majority of participants reported enjoying interacting more than observing, boredom may have affected performance to a greater degree when Observing than Interacting. 
This highlights a potential benefit of iMD-XR over standard trajectory playback: iMD-XR may be more engaging for students and researchers, enabling them to understand their molecular systems more quickly, enjoyably and effectively.

\subsection*{Comparison of iMD-XR and visuomotor incongruence}

% To distinguish this phenomenon from visuo-haptic illusions arising through visuomotor incongruence, Roebuck Williams et al.~\cite{roebuck_williams_subtle_2020} termed it `Subtle Sensing'.

% Peception of *intrinsic* properties
% Perception is shared, application to multi-person XR, multi-person interaction
There are several key differences between the Subtle Sensing approach taken within iMD-XR and pseudo-haptic systems that employ visuomotor incongruence. 
Firstly, these types of pseudo-haptic systems rely on deliberately offsetting the positions of a participant’s virtual body from their physical one to induce haptic-like sensations through cross-modal interaction. 
Consequently, in a fully-virtual VR environment the ability to experience these sensations depends on whether a participant directly interacts with the virtual object. 
Taking pseudo-haptic weight as an example, when one VR participant picks up a cube and their virtual hand moves more slowly than their physical hand, they may experience a sensory illusion of weight. 
This illusion can \textit{only be experienced by the interacting participant}; another VR participant observing close-by would simply see the cube being lifted.
By contrast, we have shown that iMD-XR participants can perceive dynamic properties of virtual molecular objects without visuomotor incongruence. 
As a consequence, \textit{non-interacting participants can also perceive these properties} within a shared virtual environment, as demonstrated here by the ability of both interactors and observers to discriminate rigidity differences.
In this sense, the perceived property is an objective characteristic of the virtual object rather than a sensory illusion that arises when interacting with the object.
This interpretation aligns with Chalmers' concept of virtual realism \cite{chalmers_reality_2022}, in which virtual objects are considered real digital entities whose properties can give rise to genuine and meaningful perceptual experiences.

A benefit to the Subtle Sensing approach taken in iMD-XR is that the field-based interaction paradigm allows multiple participants to apply forces to the same virtual object simultaneously. 
Such multi-user interaction may be more difficult to realise in pseudo-haptic systems, where virtual objects are often controlled by a single participant. 
% Together, these findings highlight the potential of the Subtle Sensing approach for multi-person, collaborative XR applications.

% The fact that perception does not rely directly on either (a) conflicting sensory information and (b) subjective experience of the resulting sensory illusions.

% changes who is able to perceive molecular properties within a virtual environment. 
% A further distinction concerns how perceptual information is conveyed, and therefore who has access to it within a virtual environment. 
% Where visuomotor incongruence is experienced only by the interactor, iMD-XR enables the properties of virtual objects to be perceived by any participant within a shared virtual environment. 

% Suitability to AR
A further consideration is the suitability of these interaction paradigms for mixed reality (MR). 
Introducing visuomotor incongruence in MR settings---where participants can simultaneously see both the physical and virtual environments---would likely reduce the effectiveness of visuo-haptic illusions because the true position of the physical hand remains visible \cite{lecuyer_simulating_2009} (to our best knowledge this setup has not yet been investigated).
By contrast, the Subtle Sensing approach is compatible with both fully virtual and mixed reality environments. 
This flexibility is increasingly important as MR technologies become more widespread, for example as provided via the passthrough functionality on devices such as the Meta Quest headsets.
Interestingly, observers in pseudo-haptic MR environments may be able to experience visuo-haptic illusions, since they can see the positions of other participants' physical and virtual hands. 
However, to our knowledge, this possibility has not yet been investigated.

% Bounding of pseudo-haptic approaches
Techniques based on visuomotor incongruence are bounded by users’ perceptual thresholds: offsets that are too small will fail to induce the desired visuo-haptic illusion, and offsets that are too large will reduce the effectiveness of the illusion by disrupting the participant's experience.
Perception of the properties of virtual molecular objects is similarly constrained at the lower end, as small differences may not be detectable. 
However, we hypothesise that this perception is not constrained at the upper end, because no sensory discrepancies are introduced to disrupt the experience.

% Visuo-haptic illusions
Although the present work demonstrates that Subtle Sensing does not depend on visuo-haptic illusions induced through visuomotor incongruence, some participants nevertheless reported experiences consistent with such illusions during interaction. 
Further investigation is required to understand the relationship between these subjective experiences and the perceptual processes underlying Subtle Sensing.

\subsection*{Applications of Subtle Sensing}

\subsubsection*{Molecular research and education}

Our findings highlight the potential of iMD-XR as a multisensory, collaborative tool for molecular research and education.
For example, groups of researchers could use iMD-XR to collectively induce complex conformational changes that would be difficult for a single participant to achieve alone. 
Bennie et al.~\cite{bennie_teaching_2019} reported improved perceived learning outcomes and engagement when using iMD-XR to teach enzyme catalysis in single-student settings. 
Similarly, Ferrell et al.~\cite{ferrell_chemical_2019} demonstrated the use of iMD-XR in an organic chemistry course for improving motivation and learning.
Such approaches could be extended to multi-student environments, for example through instructor-led demonstrations and collaborative problem-solving exercises. 
In particular, our finding that interaction improves subsequent performance when observing-only may inform the design of these sessions to enhance learning outcomes.
% Direct interaction with molecular simulations could also help students to develop an embodied understanding of the mathematical principles underlying molecular dynamics, such as how changes in angle force constants influence molecular motion.
% As such, iMD-XR has the potential to increase motivation and enjoyment, and to support more intuitive and creative approaches to complex chemical problems within research and academia.
% Such approaches could increase motivation and enjoyment, and to support more intuitive and creative approaches to complex chemical problems within research and academia.

The compatibility of iMD-XR with MR environments can help to reduce barriers to its adoption in computational chemistry. 
In an MR setting, a chemist could interact with a molecular system while simultaneously viewing associated data in a Jupyter notebook on a computer monitor, all while wearing an XR headset.

\subsubsection*{XR system design}

The field-based interaction paradigm employed in the Subtle Sensing iMD-XR approach may also be applicable beyond molecular systems. 
For example, it could support collaboration in multi-person XR games and social environments by enabling all participants to interact with virtual objects and perceive their properties. 
However, implementing field-based interaction presents practical challenges, as it may be more complex than conventional rigid body physics approaches.

\subsection*{Limitations and future work}
\label{sec:future_work}

Due to practical constraints, the present studies employed a relatively small number of trials per participant, which reduces the reliability of the estimated JNDs.
% Furthermore, two participants achieved \SI{100}{\percent} accuracy in the Interacting trials in Study One, indicating that the sampling did not capture the region of their psychometric curves corresponding to the JND.
Several participants---particularly those with chemistry backgrounds---expressed interest in taking part in follow-up experiments, suggesting that repeat participation may be a feasible approach for improving the sampling in future studies.
Furthermore, sampling efficiency may also be improved by employing an adaptive psychophysics procedure.
% Such procedures are compatible with live simulations, as used in the Interacting condition, because simulation XML files can be generated automatically.
% However, adaptive procedures would be difficult to implement with pre-recorded trajectories, as used in the Observing condition, since recordings must be prepared in advance.
% One solution could be to use AI-trained models to interact with live simulations using the approach proposed by Dhouioui et al \cite{dhouioui_ai-guided_2025}. 

In future studies it will be valuable to investigate how participant demographics influence perception of molecular properties. 
For example, we did not observe a clear relationship between chemistry experience and performance. 
This may be a result of the fact that chemistry education does not typically provide chemists with training in interactive molecular dynamics. 
As such, chemists do not have strong priors for what molecules `feel' like. 
In this respect, their starting point is comparable to non-chemists for the sorts of subtle sensing experiments described in this article. 
That said, the majority of participants in these studies did not work in chemistry or a related discipline. 
Further studies involving a larger population of chemists would be required to determine whether prior chemistry experience influences perception of molecular properties.
Furthermore, we observed that participants with less familiarity with XR and related technologies, particularly older participants who reported infrequent use of such technologies, tended to require more time to become comfortable with the XR controls and appeared to find the 2AFC task more challenging.
This observation suggests that familiarity with XR technology should be considered when recruiting participants for future studies.

Our findings show that interaction enhanced subsequent performance in the observing-only condition; however, it is unknown whether this finding extends beyond purely iMD-XR workflows.
% ---i.e. whether interaction improves the interpretation of standard, non-interactive molecular trajectories. 
% If iMD-XR improves the perception of molecular properties when observing the interactions of another person, then experience gained through interaction may also help chemists to interpret molecular motions during standard, non-interactive simulation trajectories.
Clarifying this could help to determine whether iMD-XR can improve perception during not only \textit{interactive} exploration of molecular systems, but also in the interpretation of molecular motions during standard, non-interactive simulation trajectories.
Further investigation is also required to examine how this perception generalises to different properties and other molecular systems.

% also analysis within standard computational chemistry workflows involving non-interactive molecular trajectories.

% In addition, we used the measured JNDs to predict the ability of participants to discriminate buckminsterfullerene molecules defined with the angle force constants in an alanine polypeptide system.
% Further studies are required to test this prediction and to assess how well our estimated JNDs generalise to perception of other molecular systems.
% This experimental approach could also be extended to investigate perception of other molecular properties, including other bonded properties such as bond force constants, and nonbonding interactions such as hydrogen bonds or steric clashes.

% % Effect of priming
% Another open question concerns the role of priming. 
% Pusch et al. \cite{pusch_pseudo-haptics_2011} suggest that informing participants about potential pseudo-haptic effects may enhance their perceptual experience. 
% In the present studies, participants were encouraged by the experimenter to \textit{feel} the softness of the buckyballs as well as visually observe it. 
% It remains unclear whether this guidance influenced performance, and future work could examine whether directing attention toward such sensations improves participants’ ability to interpret molecular behaviour in iMD-XR.
% Understanding this could help to inform the design of iMD-XR software more generally.

% Visuohaptic illusions - how common is this experience and does it correlate with performance?
Finally, further research is needed to better understand the phenomenological experience of interacting with these types of flexible virtual objects, including the extent to which participants experience visuo-haptic sensations. 
In particular, it would be valuable to determine whether subjective reports of visuo-haptic sensations correlate with performance in the psychophysical tasks presented here. 
Understanding this relationship could inform the design of XR systems with the aim of more effectively supporting such perceptual experiences.

% --------------------------------------------------- %
\section*{Conclusion}

In this article, we investigated `Subtle Sensing'---a phenomenon whereby XR participants perceive rigidity differences of flexible virtual objects without physical feedback---to explore more deeply what it means to `feel' purely virtual objects in XR.
Participants applied forces to dynamic molecular objects using a field-based interaction paradigm implemented through interactive molecular dynamics in extended reality (iMD-XR).
Using a two-alternative forced choice (2AFC) psychophysics approach, we quantified participants' ability to discriminate the rigidities of virtual C$_{60}$ molecules when (a) interacting directly with them and (b) observing the interactions of an iMD-XR expert.
While both interactors and observers could perceive rigidity differences, direct interaction facilitated significantly better performance than observation alone, enabling participants to perceive rigidity differences of \SI{11.5}{\percent}, compared to \SI{18.5}{\percent} for observation-only.
Furthermore, participants who undertook interaction first were better able to distinguish rigidity differences in the subsequent observation-only condition than those observing without prior experience. 

Together, these findings demonstrate that participants can perceive dynamic properties of flexible virtual objects in XR without physical feedback.
The results highlight the potential of iMD-XR for molecular research and education, and illustrate a novel approach for conveying the dynamic properties of flexible virtual objects in XR.

\section*{Technical details}
\label{sec:technical_details}

The SubtleGame source repository is provided in a data archive (\url{https://doi.org/10.5281/zenodo.19005576}).
This contains the scripts, inputs and instructions required to run the application, as well as instructions for installing the latest version of NanoVer used in this work ($0.1.2811$).
The SubtleGame source repository is also publicly available on GitHub (\url{https://github.com/IRL2/SubtleGame}).
Molecular simulation parameters and details about how the simulation data were processed are provided in \nameref{S2_Text} and \nameref{S3_Text}.

% User interactions were defined by a spherical Gaussian field potential, where the vector of forces $\mathbf{F}$ applied to an atom is given by:
% \begin{equation} \label{eq:gaussian_potential}
%     \mathbf{F} _\text{user}
%     = 
%     - c \, m \, \mathbf{d} \, e ^{- \frac{\vert \mathbf{d} \vert ^2}{ 2 } },
% \end{equation}
% where $c$ is an arbitrary scaling factor, $m$ is the atomic mass and $\textbf{d}$ is the vector between the participant's controller and the target atom.
% \Cref{fig:spherical_gaussian_potential} illustrates the magnitude of this force with respect to distance between the participant's XR controller and the target atom.

\begin{figure}
    \centering
    \includegraphics[width=0.7\linewidth]{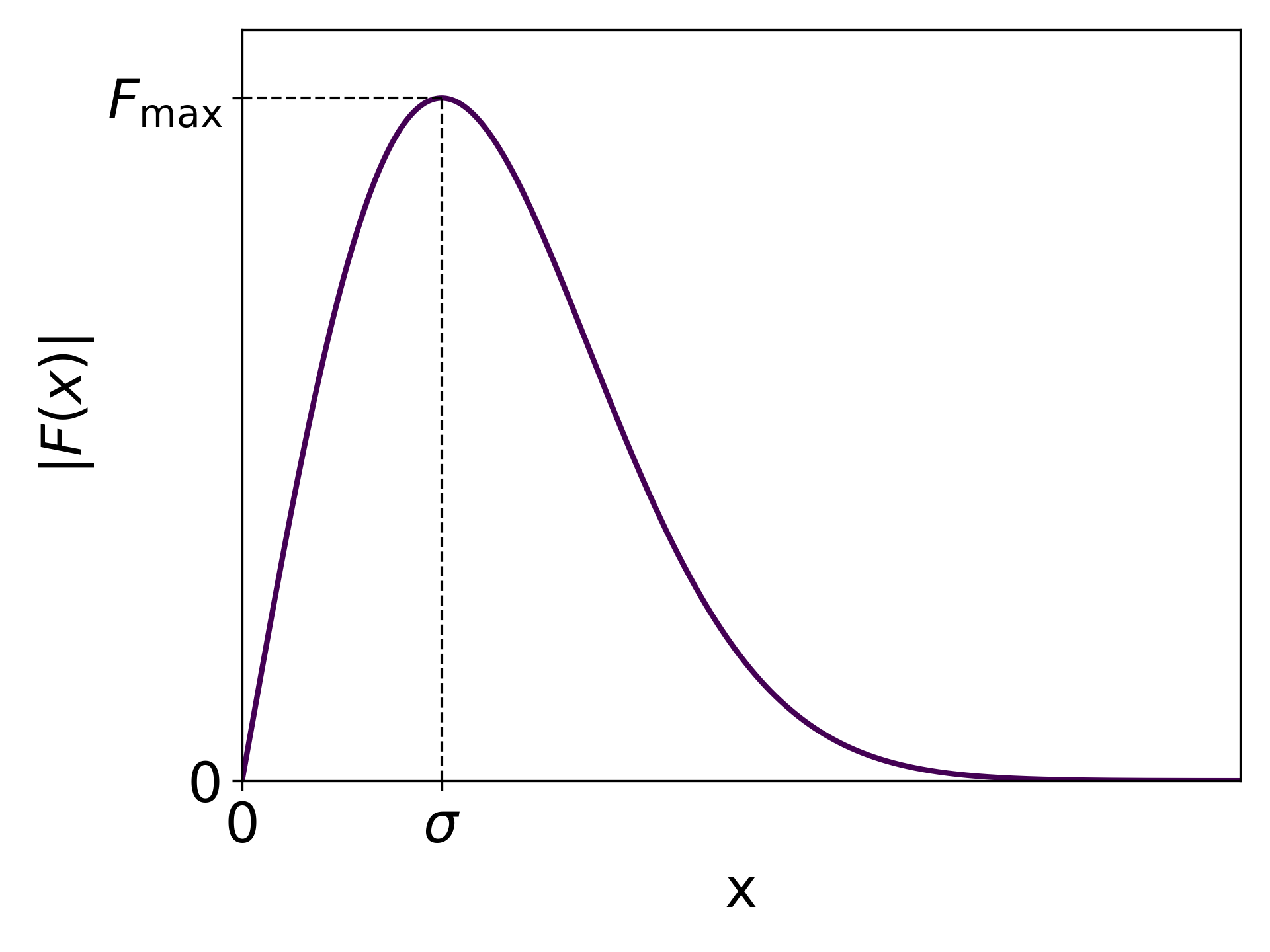}
    \vspace{0.4em}
    \caption{
    \textbf{The magnitude of the interaction force as defined by a spherical Gaussian potential in iMD-XR.}
    The variable $x$ denotes the distance between the participant's XR controller and the target atom (\Cref{eq:gaussian_potential}).
    }
    \label{fig:spherical_gaussian_potential}
\end{figure}

User interactions were defined by a spherical Gaussian field potential.
In this paradigm the vector of forces $\mathbf{F} _{\text{user,}\,i\rightarrow j} (\mathbf{q})$ applied to an atom of index $i$ by a participant's XR controller of index $j$ is given by:
\begin{equation} \label{eq:gaussian_potential}
    \mathbf{F} _{\text{user,}\,i\rightarrow j} (\mathbf{q})
    = 
    - \frac{c \times m_i}{\sigma^2} \Big( \textbf{q}_i - \textbf{g} _j \Big) e ^{- \frac{\vert \textbf{q}_i - \textbf{g} _j \vert^2}{ 2 \sigma^2} },
\end{equation}
where $c$ is an arbitrary scaling factor, $\sigma$ controls the width of the interactive field (with a default value of \SI{1}{\nano\meter}), $m_i$ is the atomic mass, $\textbf{q}_i$ is the atomic position, and $\textbf{g}_j$ is the position of the XR controller.
The magnitude of the associated force with respect to distance between the participant's XR controller and the interacted atom is illustrated in \Cref{fig:spherical_gaussian_potential}.

A scaling factor $c=400$ was used in \Cref{eq:gaussian_potential} for the 2AFC trials.
One metre in physical space mapped to three nanometres in simulation space, except for the Knot-tying task, where one metre mapped to six nanometres.
A ball-and-stick representation was used for all simulations.

% --------------------------------------------------- %
\section*{Acknowledgments}
\paragraph*{Funding.}
This project was supported by the Axencia Galega de Innovación through the Oportunius Program, the Xunta de Galicia (Research Center of Galicia accreditation 2024-2027 ED431G-2023/04) and the European Union (European Regional Development Fund - ERDF), as well as by the European Research Council under the European Union’s Horizon 2020 Research and Innovation Programme through Consolidator Grant NANOVR 866559.
\paragraph*{Author contributions.} 
All authors contributed to drafting and editing the manuscript.
RRW led the design and development of the SubtleGame software, designed and conducted the user studies, performed the data analysis, and led the writing of the manuscript.
HJS assisted with designing the user studies, creating the recordings used in SubtleGame and analysing the data.
LT contributed to the design of SubtleGame, including the game mechanics and user interface, and assisted with user studies and creating figures for the manuscript.
MW provided technical support during the development of SubtleGame.
DRG was the principal investigator and obtained funding for the project.
\paragraph*{Thanks.} We would like to thank the participants of our studies, as well as the collaborators who supported the organisation of the studies including 
Juan J. Nogueira and Manuel Alcam\'{i} at the Universidad Aut\'{o}noma de Madrid, and Rebeca Garc\'{i}a Fandi\~{n}o at the Universidade de Santiago de Compostela.
We would also like to thank the developers of NanoVer, OpenMM and the other open-source scientific software that have made this work possible.

\nolinenumbers

% Please compile your BiBTeX database using the "plos2025.bst" BibTeX style.
% This file is part of the current package.
% A sample BibTeX file is also included as "plos_bibtex_sample.bib".
%
% or
%
% Type in your references following Vancouver style and reference formatting instructions
% available at https://journals.plos.org/plosone/s/submission-guidelines#loc-references
% \begin{thebibliography}{}
% Text
% \end{thebibliography}

\bibliography{plos_bibtex_sample}

% --------------------------------------------------- %
\section*{Supporting information}

% Include only the SI item label in the paragraph heading. Use the \nameref{label} command to cite SI items in the text.

\paragraph*{S1 Table.}
\label{S1_Table}
\textbf{Locations \& dates of recruitment and participant compensation for Studies One and Two.}

\paragraph{S1 Text.}
\label{S1_Text}
\textbf{Further demographic information.}

\paragraph*{S1 File.}
\label{S1_File}
    \textbf{Study One information sheet.}

\paragraph*{S2 File.} 
\label{S2_File}
\textbf{Study Two information sheet.}

\paragraph*{S2 Text.} 
\label{S2_Text}
\textbf{Molecular simulation parameters.}

\paragraph*{S3 Text.} 
\label{S3_Text}
\textbf{Recording and processing the simulation data.}

\paragraph*{S4 Text.} 
\label{S4_Text}
\textbf{Automated procedure for quantifying buckyball deformation across rigidity scaling factors.}

\paragraph{S2 Table.}
\label{S2_Table}
\textbf{The counterbalanced factors of the recordings used in the Observing condition.}

\paragraph*{S1 Video.}
\label{S1_Video}
\textbf{Recording 1 for $x=1.036$.}

\paragraph*{S2 Video.}
\label{S2_Video}
\textbf{Recording 1 for $x=1.112$.}

\paragraph*{S3 Video.}
\label{S3_Video}
\textbf{Recording 1 for $x=1.193$.}

\paragraph*{S4 Video.}
\label{S4_Video}
\textbf{Recording 1 for $x=1.281$.}

\paragraph*{S5 Video.}
\label{S5_Video}
\textbf{Recording 1 for $x=1.375$.}

\paragraph*{S6 Video.}
\label{S6_Video}
\textbf{Recording 1 for $x=1.7$ (training).}

\paragraph{S7 Video.}
\label{S7_video}
\textbf{Example 2AFC trial for the Interacting condition.}

\paragraph*{S1 Spreadsheet.}
\label{S1_Spreadsheet}
\textbf{2AFC scores for Study One.}

\paragraph*{S2 Spreadsheet.}
\label{S2_Spreadsheet}
\textbf{2AFC scores for Study Two.}

\paragraph*{S5 Text.} 
\label{S5_Text}
\textbf{Statistical analysis details.}

\paragraph*{S3 Table.}
\label{S3_Table}
\textbf{Total score distributions for Study One.}

\paragraph*{S4 Table.}
\label{S4_Table}
\textbf{Normality test results of the total score distributions for Study One.}
Results of the normality tests using \texttt{scipy.stats.normaltest} (D'Agostino--Pearson) and \texttt{scipy.stats.shapiro} (Shapiro--Wilk). 

\paragraph*{S1 Equation.}
\label{S1_Equation}
\textbf{Cohen's \textit{d} for independent samples.}
Used to calculate the standardised effect size in units of number of standard deviations for independent distributions.

\paragraph*{S2 Equation.}
\label{S2_Equation}
\textbf{Cohen's \textit{d} for paired samples.}
Used to calculate the standardised effect size in units of number of standard deviations for paired distributions.

\paragraph{S6 Text.}
\label{S6_Text}
\textbf{Comments on outlier in 2AFC total scores.}

\paragraph*{S7 Text.}
\label{S7_Text}
\textbf{Psychometric analysis details.}

\paragraph{S8 Text.}
\label{S8_Text}
\textbf{Discussion about unusually low average probability for $x = 1.036$ shown on the psychometric curve.}

\paragraph*{S1 Figure.}
\label{S1_Figure}
\textbf{Total RMSD, per-atom RMSDs of the interacted atoms and user force magnitudes during Trial 14 (3 of 5 for $x=1.036$) from P1.} 

\paragraph*{S2 Figure.}
\label{S2_Figure}
\textbf{Total RMSD, per-atom RMSDs of the interacted atoms and user force magnitudes during Trial 4 (1 of 5 for $x=1.036$) from P11.} 

\paragraph*{S3 Figure.}
\label{S3_Figure}
\textbf{Total RMSD, per-atom RMSDs of the interacted atoms and user force magnitudes during Trial 22 (5 of 5 for $x=1.036$) from P9.} 

\paragraph*{S4 Figure.}
\label{S4_Figure}
\textbf{Total RMSD, per-atom RMSDs of the interacted atoms and user force magnitudes during Trial 8 (3 of 5 for $x=1.036$) from P24.} 

\paragraph*{S5 Figure.}
\label{S5_Figure}
\textbf{Total RMSD, per-atom RMSDs of the interacted atoms and user force magnitudes during Recording 1 for $x=1.036$ in Study One.} 

\paragraph*{S6 Figure.}
\label{S6_Figure}
\textbf{Total RMSD, per-atom RMSDs of the interacted atoms and user force magnitudes during Recording 2 for $x=1.036$ in Study One.} 

\paragraph*{S7 Figure.}
\label{S7_Figure}
\textbf{Total RMSD, per-atom RMSDs of the interacted atoms and user force magnitudes during Recording 3 for $x=1.036$ in Study One.} 

\paragraph*{S8 Figure.}
\label{S8_Figure}
\textbf{Total RMSD, per-atom RMSDs of the interacted atoms and user force magnitudes during Recording 4 for $x=1.036$ in Study One.} 

\paragraph*{S9 Figure.}
\label{S9_Figure}
\textbf{Total RMSD, per-atom RMSDs of the interacted atoms and user force magnitudes during Recording 1 for $x=1.375$ in Study One.} 

\paragraph*{S10 Figure.}
\label{S10_Figure}
\textbf{Total RMSD, per-atom RMSDs of the interacted atoms and user force magnitudes during Recording 2 for $x=1.375$ in Study One.} 

\paragraph*{S11 Figure.}
\label{S11_Figure}
\textbf{Total RMSD, per-atom RMSDs of the interacted atoms and user force magnitudes during Recording 3 for $x=1.375$ in Study One.} 

\paragraph*{S12 Figure.}
\label{S12_Figure}
\textbf{Total RMSD, per-atom RMSDs of the interacted atoms and user force magnitudes during Recording 4 for $x=1.375$ in Study One.} 

\paragraph*{S13 Figure.}
\label{S13_Figure}
\textbf{Percentage of correct responses per simulation per scaling factor for the Interacting trials in Study One.}

\paragraph*{S14 Figure.}
\label{S14_Figure}
\textbf{Percentage of correct responses per simulation per scaling factor for the Observing trials in Study One.}

\paragraph*{S3 File.} 
\label{S3_File}
\textbf{Study One questionnaire.}

\paragraph*{S4 File.} 
\label{S4_File}
\textbf{Study Two questionnaire}

\paragraph*{S3 Equation.}
\label{S3_Equation}
\textbf{Calculating the relative difference between two angle force constants as a Weber fraction.}

% \paragraph*{S5 Table.} 
% \label{S5_Table}
% \textbf{Weber fractions of the 17-alanine angle force constants.}

\paragraph*{S4 Equation.}
\label{S4_Equation}
\textbf{Converting the JND for decreasing rigidity from Study Two into the predicted threshold for increasing rigidity.}

\paragraph*{S9 Text.}
\label{S9_Text}
\textbf{Themes from thematic analysis of questionnaire responses for Study One.}

\end{document}